\documentclass[]{aastex631}
\usepackage{CJK}

\begin{document}
\begin{CJK*}{UTF8}{gbsn}
\title{The Stellar Abundances and Galactic Evolution Survey (SAGES). IV. Surface Gravity Estimation and Giant-Dwarf Separation with the DDO51 Filter}

\author[0000-0003-0278-7137]{Qiqian Zhang}
\affiliation{National Astronomical Observatories, Chinese Academy of Sciences, Beijing 100101,  People's Republic of China;  gzhao@nao.cas.cn; zfan@bao.ac.cn}
\affiliation{ School of Astronomy and Space Science, University of Chinese Academy of Sciences, Beijing 100049, People's Republic of China
}

\author[0000-0002-6790-2397]{Zhou Fan}
\affiliation{National Astronomical Observatories, Chinese Academy of Sciences, Beijing 100101,  People's Republic of China;  gzhao@nao.cas.cn; zfan@bao.ac.cn}
\affiliation{ School of Astronomy and Space Science, University of Chinese Academy of Sciences, Beijing 100049, People's Republic of China
}

\author[0000-0002-8980-945X]{Gang Zhao}

\affiliation{National Astronomical Observatories, Chinese Academy of Sciences, Beijing 100101,  People's Republic of China;  gzhao@nao.cas.cn; zfan@bao.ac.cn}
\affiliation{ School of Astronomy and Space Science, University of Chinese Academy of Sciences, Beijing 100049, People's Republic of China
}

\author[0000-0002-3257-9286]{Ying Wu}
\affiliation{National Astronomical Observatories, Chinese Academy of Sciences, Beijing 100101,  People's Republic of China;  gzhao@nao.cas.cn; zfan@bao.ac.cn}

\author[0000-0002-9702-4441]{Wei Wang}
\affiliation{National Astronomical Observatories, Chinese Academy of Sciences, Beijing 100101,  People's Republic of China;  gzhao@nao.cas.cn; zfan@bao.ac.cn}

\author[0000-0001-8424-1079]{Kai Xiao}
\affiliation{ School of Astronomy and Space Science, University of Chinese Academy of Sciences, Beijing 100049, People's Republic of China
}

\author[0009-0007-5610-6495]{Hongrui Gu}
\affiliation{National Astronomical Observatories, Chinese Academy of Sciences, Beijing 100101,  People's Republic of China;  gzhao@nao.cas.cn; zfan@bao.ac.cn}
\affiliation{ School of Astronomy and Space Science, University of Chinese Academy of Sciences, Beijing 100049, People's Republic of China
}
\author[0000-0001-6637-6973]{Jie Zheng}
\affiliation{National Astronomical Observatories, Chinese Academy of Sciences, Beijing 100101,  People's Republic of China;  gzhao@nao.cas.cn; zfan@bao.ac.cn}

\author[0000-0003-2868-8276]{Jingkun Zhao}
\affiliation{National Astronomical Observatories, Chinese Academy of Sciences, Beijing 100101,  People's Republic of China;  gzhao@nao.cas.cn; zfan@bao.ac.cn}
\author[0009-0000-4835-7525]{Chun Li}
\affiliation{National Astronomical Observatories, Chinese Academy of Sciences, Beijing 100101,  People's Republic of China;  gzhao@nao.cas.cn; zfan@bao.ac.cn}
\author[0000-0002-8442-901X]{Yuqin Chen}
\affiliation{National Astronomical Observatories, Chinese Academy of Sciences, Beijing 100101,  People's Republic of China;  gzhao@nao.cas.cn; zfan@bao.ac.cn}

\author[0000-0003-2471-2363]{Haibo Yuan}
\affiliation{Department of Astronomy, Beijing Normal University, Beijing 100875, People's Republic of China}
\author[0000-0002-0389-9264]{Haining Li}
\affiliation{National Astronomical Observatories, Chinese Academy of Sciences, Beijing 100101,  People's Republic of China;  gzhao@nao.cas.cn; zfan@bao.ac.cn}
\author[0000-0003-0173-6397]{Kefeng Tan}
\affiliation{National Astronomical Observatories, Chinese Academy of Sciences, Beijing 100101,  People's Republic of China;  gzhao@nao.cas.cn; zfan@bao.ac.cn}
\author[0000-0001-7255-5003]{Yihan Song}
\affiliation{National Astronomical Observatories, Chinese Academy of Sciences, Beijing 100101,  People's Republic of China;  gzhao@nao.cas.cn; zfan@bao.ac.cn}
\author[0000-0001-7865-2648]{Ali Luo}
\affiliation{National Astronomical Observatories, Chinese Academy of Sciences, Beijing 100101,  People's Republic of China;  gzhao@nao.cas.cn; zfan@bao.ac.cn}
\author{Nan Song}
\affiliation{China Science and Technology Museum, Beijing 100101, People's Republic of China}
\author[0009-0008-3430-1027]{Yujuan Liu}
\affiliation{National Astronomical Observatories, Chinese Academy of Sciences, Beijing 100101,  People's Republic of China;  gzhao@nao.cas.cn; zfan@bao.ac.cn}
\author[0000-0002-8337-4117]{Yaqian Wu}
\affiliation{National Astronomical Observatories, Chinese Academy of Sciences, Beijing 100101,  People's Republic of China;  gzhao@nao.cas.cn; zfan@bao.ac.cn}



\begin{abstract}

Reliable estimation of stellar surface gravity (log\,$g$) for a large sample is crucial for evaluating stellar evolution models and understanding galactic structure; However, it is not easy to accomplish due to the difficulty in gathering a large spectroscopic data set. Photometric sky survey using a specific filter, on the other hand, can play a substantial role in the assessment of log\,$g$. The Stellar Abundances and Galactic Evolution Survey (SAGES) utilizes eight filters to provide accurate stellar parameters for $\sim10^{7}$ stars, with its DDO51 intermediate-band filter specifically designed for robust log\,$g$ determination. In this work, the observed SAGES $u_{\rm SC}$ and $v_{\rm SAGES}$ photometry, the synthetic photometry in $g$, $r$, $i$, and DDO51 bands derived from \textit{Gaia} XP spectra are employed to investigate the importance of the DDO51 filter in the determination of log\,$g$. We applied machine-learning-based extinction correction and employed XGBoost models, trained on stellar parameters from LAMOST, to predict log\,$g$ using photometric data. By comparing model predicted log\,$g$ with LAMOST values, we find that including DDO51 filter improve the accuracies of log\,$g$ estimates by 21.0\%  (from 0.224\,dex to 0.177\,dex) overall, and by 26.5\% (from 0.302\,dex to 0.222\,dex ) for GK-type stars, as compared to those obtained without DDO51. The DDO51 filter is also validated to be particularly effective for metal-poor stars ([Fe/H]$<$-1.0), where it significantly mitigates systematic biases. Our findings highlight the diagnostic power of the SAGES DDO51 filter, providing enhanced stellar characterization vital for future in-depth studies of the Milky Way.

\end{abstract}



\section{Introduction} \label{sec:intro}
Large samples of stellar parameters are essential for understanding the formation, evolution, dynamical structure, and distribution of the stellar population of the Milky Way. Stellar parameters can be described by nine dimensions\footnote{Including three spatial coordinates, three kinematic parameters, and the three primary atmospheric parameters (effective temperature,$T_{\rm eff}$; metallicity,[M/H]; and surface gravity, log\,$g$).}\citep{ivezic2008milky}, while the latest release of the \textit{Gaia} DR3 (\citealt{vallenari2023Gaia}) provides the positions and kinematics of more than one billion stars, but the relatively shallow limiting magnitude of \textit{Gaia} spectra makes it difficult to derive atmospheric parameters with sufficient precision for faint stars.

However, high-resolution spectroscopic observations require long integration times to reach a sufficient signal-to-noise ratio (SNR) for stellar studies, making spectroscopic surveys for stellar parameter determination very time-consuming and limited in depth. On the other hand, multi-band photometry offers a more efficient means to estimate stellar atmospheric parameters. Although photometry lacks the high spectral resolution necessary to resolve individual spectral lines as spectroscopy does, it still provides valuable information about stellar parameters through systematic variations in color indices. By carefully choosing photometric systems, specific stellar colors can be measured and used as proxies for atmospheric parameters using empirically calibrated color-parameter relationships, a process achievable with exposure times of only a few minutes \citep{casagrande2014synthetic}. Photometry also allows custom filter designs that target specific spectral features, enhancing sensitivity to atmospheric parameters while mitigating interference from other spectral regions \citep{rix2022poor}. 

However, photometry faces inherent challenges that must be addressed for accurate parameter estimation. For example, interstellar extinction can redden starlight \citep{cardelli1989relationship}, and different combinations of atmospheric parameters can produce similar colors, leading to degeneracies and uncertainties in parameter estimation.
A particularly significant challenge arises from the distinct atmospheric structures (especially surface gravity) of giant and dwarf stars. These differences often require the development of separate models for each type to accurately derive atmospheric parameters. Failing to distinguish between these populations can lead to misclassification (e.g., misclassifying foreground dwarfs for distant giants) and substantial errors, especially in metallicity analysis and stellar population studies \citep{geisler1984luminosity}. Therefore, robust classification of giants and dwarfs, or more generally, accurate surface gravity determination, is critical for leveraging photometric data effectively and achieving high precision in parameter estimation.

To address these challenges, various photometric surveys have been developed with tailored scientific goals \citep{bessell2005standard}. One such survey is the Stellar Abundances and Galactic Evolution Survey (SAGES; \citealt{wang2013stromgren,zheng2018sage,Fan2018sage,zheng2019sage}), a large-scale multi-band photometric survey covering 12,000 square degrees of the northern sky. It employs a photometric system specifically designed to be sensitive to stellar atmospheric parameters.

Previous studies have attempted to classify stars using photometric and parallax-based approaches. \citet{huang2023beyond} derived atmospheric parameters of 26 million stars using SAGES DR1 catalog. Due to the lack of photometric data sensitive to stellar surface gravity, they relied on parallaxes and \textit{Gaia} luminosities to empirically separate dwarfs and giants, and then developed separate polynomial fitting models for each group. However, a large fraction of the fainter objects in the \textit{Gaia} catalog lack parallax information with sufficient SNR, limiting the reliable sample size. Furthermore, the study did not derive surface gravity estimates, which are essential for refining stellar classifications. Similarly, \citet{gu2025stellar} applied machine learning methodologies to estimate atmospheric parameters for 21 million stars using SAGES DR1 data and other surveys.

\begin{figure}
    \centering
    \includegraphics[width=0.7\linewidth]{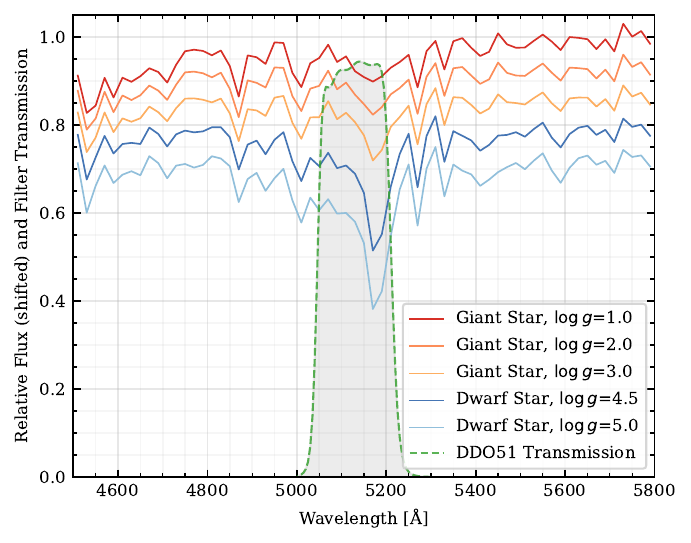}
    \caption{Schematic diagram illustrating the DDO51 filter's sensitivity to stellar surface gravity. Solid lines represent theoretical spectra of dwarf and giant stars with identical $T_{\rm eff} =5000\, {\rm K}$ and [M/H]$=-0.5$. The green dashed line indicates the DDO51 filter transmission. The spectra have been normalized and vertically shifted for clarity. The depth of the \ion{Mg}{1} absorption features within the DDO51 passband (highlighted by the gray shaded region) increases markedly with increasing log\,$g$, demonstrating the filter's effectiveness in distinguishing between different surface gravities. }
    \label{fig:DDO51}
\end{figure}

In this paper, we focus on the DDO51 filter, a medium-band filter centered at 5130 \AA (bandwidth 162 \AA), within the SAGES photometric system, which is specifically designed to measure the strength of the \ion{Mg}{1}\,b triplet ($\lambda$5167, $\lambda$5172, $\lambda$5183) in late-type stars. This spectral feature exhibits pronounced pressure-broadened wings that are highly sensitive to stellar surface gravity, even in metal-poor stars, due to the relative insensitivity of Mg I/Mg II ionization balance to non-local thermodynamic equilibrium (non-LTE) effects in F- and G-type stars \citep{fuhrmann1997surface, osorio2015mg}. 
Figure \ref{fig:DDO51} illustrates the DDO51 band and three theoretical stellar spectra with varying log\,$g$ values but the same effective temperature and metallicity. A clear trend emerges: the strength of the absorption features near 5100 \AA, primarily attributed to \ion{Mg}{1} \,b lines, increases with decreasing log\,$g$. This demonstrates the filter's intrinsic sensitivity to surface gravity, providing a physical basis for its application.

The strong correlation between this feature's strength and stellar surface gravity was first noted by \citet{ohman1936red} and \citet{thackeray1939intensity}. Subsequent studies, including \citet{clark1979photoelectric}, successfully separated dwarfs and giants with $0.8 < (B-V) < 1.35 $ using the DDO51 filter in combination with other bands. \citet{geisler1984luminosity} further demonstrated that the (M-51) color index is a powerful luminosity discriminator, capable of separating G- and K-type giants and dwarfs with a precision of up to 0.2 magnitudes. Although some confusion may occur between metal-poor dwarfs ([Fe/H] $\leq -1.0$) hotter than 4700 K and metal-rich giants ([Fe/H]$\geq +0.3$), the DDO51 filter exhibits a strong ability to distinguish metal-poor red giants and metal-rich dwarfs. \citet{casey2018infrared} combined $g-DDO51$ photometry for dwarf/giant separation with WISE infrared colors to effectively select metal-poor giant candidates in the Kepler field.

The DDO51 filter's effectiveness has also been validated in studies of open clusters \citep{majewski2000exploring} and in target selection for the APOGEE survey \citep{zasowski2013target}. However, its application has been limited to small samples or pre-selection for surveys like APOGEE. The SAGES survey represents the first use of DDO51 as a primary filter in a wide-field, deep photometric survey, aiming to refine stellar surface gravity estimates across a vast stellar population.

However, reliance on single DDO51 color-color diagrams may be insufficient due to inherent degeneracies. As shown in Appendix A using synthetic theoretical spectra, stars with different combinations of effective temperature, surface gravity, and metallicity can exhibit the same position in DDO51 color-color diagrams. By incorporating information from multiple SAGES filters, we can break these degeneracies and achieve more accurate and precise parameter estimations.

This paper investigates the effectiveness of the DDO51 filter within the SAGES photometric system for differentiating stars across a range of surface gravities. By leveraging the precise stellar atmospheric parameters provided by contemporary spectroscopic surveys, we move beyond the traditional binary classification of giant and dwarf stars. By integrating data from other SAGES filters, we endeavor to derive more precise estimates of stellar surface gravities. These parameters, in turn, enable the determination of stellar luminosities and evolutionary stages, thereby contributing to a more comprehensive understanding of Galactic structure and evolution via analysis of a large, well-characterized stellar sample.

The paper is organized as follows: Section 2 presents the data sets used in our analysis, including the generation of synthetic photometry from \textit{Gaia} XP spectra and the quality criteria applied to construct reliable samples from APOGEE DR17 and LAMOST DR10. Section 3 introduces our methodology, beginning with extinction correction. We derive DDO51 extinction coefficients via synthetic photometry and outline the XGBoost-based machine learning approach to predict stellar surface gravities from extinction-corrected magnitudes. Section 4 presents the results of extinction correction and XGboost experiments. Section 5 summarizes our findings.

\section{Data} \label{sec:data}
 Theoretical stellar spectra represent idealized stellar atmospheric conditions, but real-world observations are more complex, influenced by various factors such as model imperfections and observational errors. As demonstrated by \citet{wang2023stellar}, theoretical and observed spectra often occupy distinct regions in the high-dimensional flux space due to these discrepancies. To assess the performance of the DDO51 filter system prior to the SAGES DDO51 catalog release, we generated synthetic DDO51 magnitudes using \textit{Gaia} DR3 XP spectra \citep{vallenari2023Gaia} alongside other SAGES bands. This approach enables us to validate the filter's sensitivity to stellar parameters, thereby assessing the DDO51 filter's potential and providing a foundation for the SAGES DDO51 data reduction and subsequent data processing (e.g., zero-point calibration; Xiao et al., in prep.).

 We cross-matched this synthetic photometry with reliable stellar atmospheric parameters from APOGEE DR17 \citep{majewski2017apache} and LAMOST DR10 \citep{cui2012large} using our custom tool\footnote{\url{https://github.com/zqqian/GaiaCatalogQuery}} \citep{zhang2023efficient}, constructing a comprehensive data set combining multi-band photometry and high-quality stellar labels for our subsequent analysis, as detailed in the following sections.
 
\subsection{Photometric Data}

\begin{deluxetable}{lcccccccc}
\tablecaption{SAGES Filter Properties\label{tab:SAGES_filters}}
\tablewidth{0pt}
\tablehead{
\colhead{\textbf{Filter}} & \colhead{$u_{\rm SC}$} & \colhead{$v_{\rm SAGES}$} & \colhead{$g$} & \colhead{$r$} & \colhead{$i$} & \colhead{$H\alpha_n$} & \colhead{$H\alpha_w$} & \colhead{DDO51}
}
\startdata
\textbf{Central Wavelength (\AA)} & 3425 & 3950 & 4686 & 6166 & 7480 & 6563 & 6563 & 5130 \\
\textbf{Bandwidth (\AA)}          & 314  & 290  & 1280 & 1150 & 1230 &  29  & 136  &  162 \\
\enddata
\end{deluxetable}
\subsubsection{SAGES Data}
The SAGES survey employs a photometric system specifically designed to be sensitive to stellar atmospheric parameters, utilizing a set of specialized filters summarized in Table~\ref{tab:SAGES_filters}, including: a) the $u_{\rm SC}$ band, analogous to Str\"{o}mgren-Crawford's $u$, which covers the Balmer jump and is sensitive to stellar surface gravity of A/F-type stars; b) the SAGES-designed $v_{\rm SAGES}$ band, which covers the Ca H/K lines and is very sensitive to stellar metallicity; c) the SDSS-like $g$, $r$, and $i$ bands, which are used to derive stellar effective temperatures; d) the $H\alpha_{\text{n}}$, $H\alpha_{\text{w}}$ bands, which help estimate the interstellar extinctions; and e) the DDO51 band, sensitive to the surface gravity of late-type stars.

The $u_{\rm SC}$ and $v_{\rm SAGES}$ bands were observed using the 2.3-m Bok telescope at Kitt Peak, whereas the $g$, $r$, and $i$ bands were obtained with the Nanshan One-meter Wide-field Telescope (NOWT). Observations in the DDO51 band are currently in progress at NOWT.
The survey has already completed observations in the $u_{\rm SC}$, $v_{\rm SAGES}$, $g$, $r$, and $i$ bands and is currently observing in the DDO51 band, with completion expected by 2027. The first data release includes the DR1 catalog for the $u_{\rm SC}$ and $v_{\rm SAGES}$ bands \citep{fan2023stellar} and the DR1s catalog for the $g$, $r$, and $i$ bands \citep{li2024stellar}. Table \ref{tab:SAGES_limits} presents the magnitude limits for the released SAGES catalogs ($u_{\rm SC}$ , $v_{\rm SAGES}$, $g$, $r$, $i$) along with preliminary estimates for the ongoing DDO51 observations. We apply standard quality flags (\texttt{FLAG\_U == 0} and \texttt{FLAG\_V == 0}) to ensure data reliability when using SAGES DR1 data.

\begin{deluxetable}{lccc}
\tablecaption{Typical SAGES Photometric Depth\label{tab:SAGES_limits}}
\tablewidth{0pt}
\tablehead{
\colhead{\textbf{Filter}} & \colhead{Completeness Limit\tablenotemark{a}} & \colhead{High-Precision Limit\tablenotemark{b}} \\
\colhead{} & \colhead{(mag)} & \colhead{(mag)} 
}
\startdata
$u_{\rm SC}$    & 20.4   & 17.0 \\
$v_{\rm SAGES}$       & 20.3   & 18.0 \\ 
$g$         & 19.2   & 16.6 \\
$r$        & 19.1   & 16.5 \\
$i$         & 18.2   & 15.5 \\
DDO51      & $\sim 19$ & $\sim 16$  \\ 
$H\alpha_n$, $H\alpha_w$      & $\sim 18$ & $\sim 15$  
\enddata
\tablecomments{Typical photometric depths (median values) for SAGES filters. 
             $u_{\rm SC}$, $v_{\rm SAGES}$ data are from SAGES DR1 \citep{fan2023stellar}; 
             $g$, $r$, $i$ data are from SAGES DR1s \citep{li2024stellar}; 
             DDO51 values are preliminary estimates based on ongoing observations. $H\alpha_n$, $H\alpha_w$ values are designed values.}
\tablenotetext{a}{Approximate magnitude at the turnover of the source number counts histogram (completeness limit).}
\tablenotetext{b}{Approximate magnitude for S/N $\approx$ 100 ($\sim$0.01 mag precision).}
\end{deluxetable}

\subsubsection{Synthetic Photometry from \textit{Gaia} XP Spectra}

\textit{Gaia} DR3 provides low-resolution spectra for approximately 220 million sources, derived from combined Blue Photometer (BP) and Red Photometer (RP) prism observations (collectively termed XP spectra).
These spectra span a wavelength range of 330 to 1050 nm with a resolution of $R \approx 50-160$, intermediate between typical spectroscopic and photometric surveys. Each spectrum is calibrated and represented by 110 coefficients applied to a set of basis functions \citep{de2023Gaia}. This capability enables the recovery of absolute flux in different passbands for each source and allows synthetic photometry to be performed using XP spectra, allowing the generation of magnitudes in diverse passbands based on absolute flux and filter passbands (\citealt{montegriffo2023gaia,xiao2023j}). 
 
Several studies have already explored the direct extraction of stellar parameters from XP spectra. For instance, \citet{andrae2023Gaia} employed the General Stellar Parameterizer from Photometry (GSP-Phot) using a Bayesian forward-modeling approach to estimate atmospheric parameters, extinction, and distances from XP data. Similarly, \citet{rix2022poor} and \citet{andrae2023robust} successfully employed XGBoost machine learning algorithms to derive stellar parameters directly from the XP coefficients, finding that incorporating synthetic photometry - even synthesized from the same XP coefficients - enhanced the accuracy of parameter estimations. These results indicate that while spectra provide a continuous and information-rich representation of astrophysical sources, strategically designed photometric systems can amplify critical features for extracting specific parameters by focusing on diagnostically relevant wavelength regions in a manner that is more readily leveraged by algorithms. This highlights the inherent value and non-redundant nature of photometry, even when spectral data are available.

Given that the wavelength coverage of \textit{Gaia} XP spectra do not fully encompass the $u_{\rm SC}$ band, and studies by \citet{rix2022poor} and \citet{andrae2023robust} have indicated that the precision of XP spectra is lower at the blue end, potentially impacting the completeness of the synthesized photometry.  We utilized the observational data $u_{\rm SC}$- and $v_{\rm SAGES}$-band from SAGES DR1. For the remaining SAGES $g$, $r$, $i$, and DDO51 bands, we generated synthetic photometry using XP spectra corrected by \citet{huang2024comprehensive}, which analyzed and corrected XP spectra using CALSPEC and NGSL data, effectively reducing a significant portion of the systematic errors for sources with E(B-V) $<$ 0.8. 
As an additional quality control step to remove spurious sources and mitigate cross-matching errors, we required consistency between observed SAGES $v_{\rm SAGES}^{\rm obs}$ and synthetic XP photometry $v_{\rm SAGES}^{\rm syn}$, accepting only sources with $|v_{\rm SAGES}^{\rm obs} - v_{\rm SAGES}^{\rm syn}| < 0.25$ mag. 
Standard quality flags (\texttt{Gaia['ruwe']} $<$ 1.3 and \texttt{Gaia['ipd\_frac\_multi\_peak']} $<$ 5) are also applied to remove potential unresolved binaries and to ensure astrometric reliability of the Gaia sources.

\subsection{Stellar Parameters Data}
\subsubsection{APOGEE}
The Apache Point Observatory Galactic Evolution Experiment (APOGEE; \citealt{majewski2017apache}) is a large-scale, near-infrared spectroscopic survey of stars. APOGEE is distinguished among large spectroscopic surveys by its ability to obtain high-resolution ($R \sim 22,500$) spectra across the entire Milky Way.

The most recent data release, DR17, provides spectra, radial velocities, stellar atmospheric parameters, and individual elemental abundances for over 657,000 sources, sampling all major components of the Milky Way. We performed some steps to select suitable data for our analysis. We began by cross-matching the APOGEE DR17 catalog with the \textit{Gaia} XP and SAGES DR1 catalog using a 1-arcsecond positional tolerance. For sources with multiple APOGEE observations, we retained only the record with the highest SNR. We then applied quality cuts to exclude unreliable measurements, following the recommendations from the APOGEE documentation:

\begin{itemize}
    \setlength{\itemsep}{7pt}
    \item \texttt{ASPCAPFLAG['starbad']} == 0, to exclude systematically biased or statistically untrustworthy parameters.
    \item \texttt{ASPCAPFLAG['starwarn']} == 0, to remove parameters with potentially blurry measurements.
\end{itemize}

In addition, we applied some physical selection criteria to refine the sample further:

\begin{itemize}
    \item $|\text{GLAT}|> 10^{\circ}$, to reduce crowding and extinction in the Galactic plane, and to match the coverage of the SAGES survey.
    \item $T_{\rm{eff}} > 3750$\,K, to avoid cooler stars that are more susceptible to systematic uncertainties in APOGEE's model atmospheres.
\end{itemize}

\begin{figure}[ht!]
    \centering
    \gridline{\fig{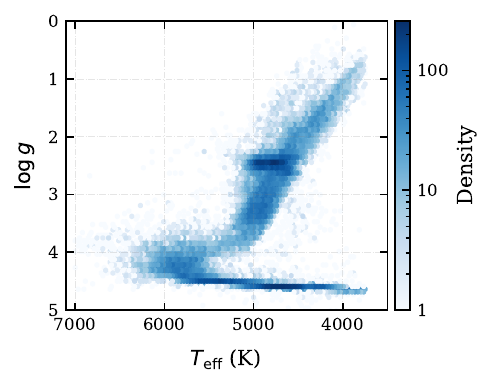}{0.32\textwidth}{(a) Kiel diagram (APOGEE)}
              \fig{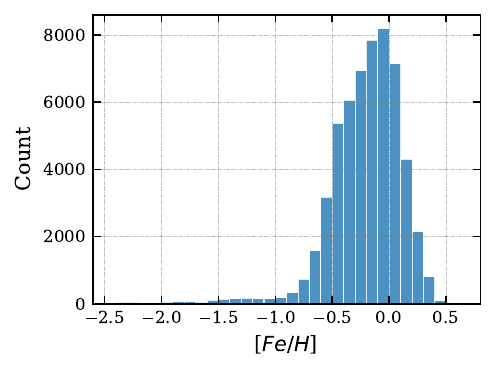}{0.32\textwidth}{(b) [Fe/H] (APOGEE)}
              \fig{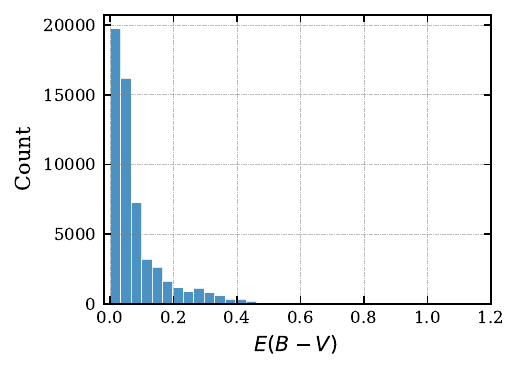}{0.32\textwidth}{(c) E(B-V) (APOGEE)}
             }
    \gridline{\fig{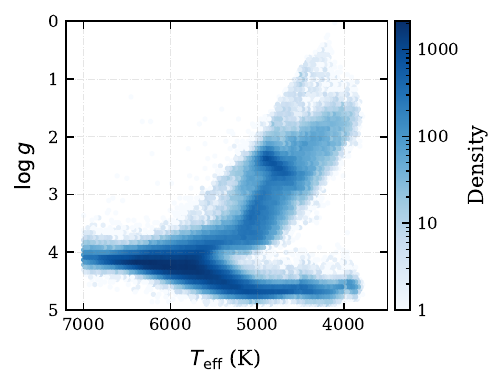}{0.32\textwidth}{(d) Kiel diagram (LAMOST)}
              \fig{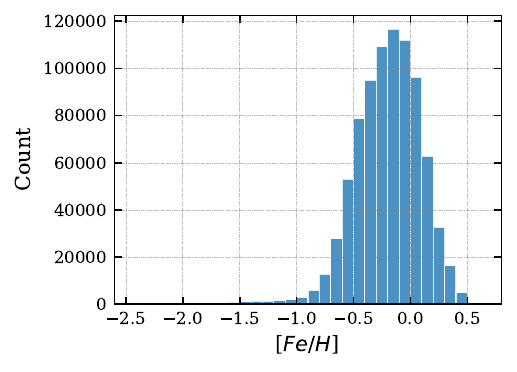}{0.32\textwidth}{(e) [Fe/H] (LAMOST)}
              \fig{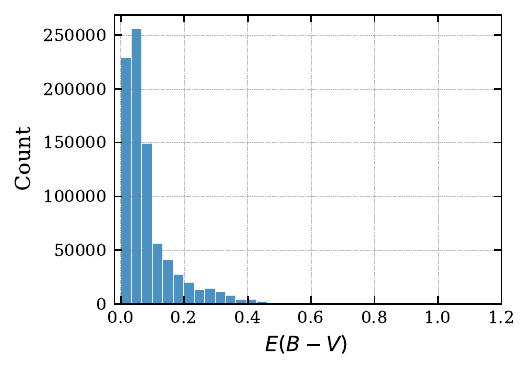}{0.32\textwidth}{(f) E(B-V) (LAMOST)}
             }
    \caption{Distributions of stellar atmospheric parameters.
    Top row (APOGEE-\textit{Gaia}-SAGES data set): (a) Kiel diagram, showing log\,$g$ vs. $T_{\rm eff}$, with color density representing the relative number of stars. (b) [Fe/H] distribution histogram. (c) E(B-V) distribution histogram.
    Bottom row (LAMOST-\textit{Gaia}-SAGES data set): (d) Kiel diagram, similar to (a). (e) [Fe/H] distribution histogram, similar to (b). (f) E(B-V) distribution histogram, similar to (c).}
    \label{fig:catalogs} 
\end{figure}

After these selections, our final sample consists of 56,141 stars. The distributions of stellar parameters for this sample are shown in top row of Figure~\ref{fig:catalogs}. This data set will subsequently be used to derive extinction coefficients and to provide an external validation for the model's predictions of log\,$g$.

\subsubsection{LAMOST}
The Large Sky Area Multi-Object Fiber Spectroscopic Telescope (LAMOST) is a special reflecting Schmidt telescope equipped with 4,000 optical fibers on its focal plane, enabling simultaneous observations of 4,000 targets within a 20 square degree field of view \citep{cui2012large}. The DR10 of the LAMOST spectroscopic survey contains over 20 million spectra. Specifically, the LAMOST Low-Resolution Survey (LRS) Stellar Parameter Catalog of A, F, G and K Stars includes stellar parameters for over seven million objects. We cross-matched the LAMOST DR10 LRS stellar parameter catalog with the \textit{Gaia} XP and the SAGES DR1 catalog. To eliminate duplicate sources, we retained, for each group of same \texttt{source\_id}, the one with the smallest positive value of \texttt{logg\_err}. Quality cuts were then applied to select reliable stellar parameters as follows:
\begin{itemize}
    \setlength{\itemsep}{7pt}
    \item 0 $<$ \texttt{LAMOST['logg\_err']} $<$ 0.1, to select stars with well-constrained surface gravity measurements.
    \item 0 $<$ \texttt{LAMOST['feh\_err']} $<$ 0.1, to ensure reliable metallicity estimates.
    \item 3750 K $< T_{\rm{eff}} <$ 7000 K, to restrict the sample to F, G, and K-type stars with model atmospheres that are well-calibrated and less prone to systematics.
    \item \(|\text{GLAT}| > 10^{\circ}\), to avoid the Galactic plane.
\end{itemize}

These criteria resulted in a final sample of 839,862 sources. The distributions of stellar parameters for this sample are shown in bottom row of Figure~\ref{fig:catalogs}.

\section{Methods}
This section details the methodology used to correct for interstellar extinction and to predict stellar surface gravity, with a particular focus on evaluating the impact of the DDO51 filter. The analysis is based on machine learning techniques, primarily utilizing the XGBoost algorithm.

\subsection{Extinction Correction and Extinction Coefficient of DDO51}\label{sec:extinction}

Giant stars, due to their higher luminosity, can be detected at much greater distances than dwarf stars, making them more susceptible to interstellar extinction and the associated reddening. Although the SAGES survey area was strategically selected at $|\text{GLAT}|> 10^{\circ}$, aims to avoid high-extinction regions in the Galactic disk, the presence of interstellar extinction can still systematically shift the positions of distant stars in color-color diagrams. These shifts can compromise the accuracy of stellar parameter estimations.
Therefore, correcting the observed data for interstellar extinction is essential to accurately recover the photometric properties of these stars and to accurately determine their surface gravity and other fundamental parameters.

The color index $H\alpha_{\text{n}}$-$H\alpha_{\text{w}}$ is largely insensitive to interstellar extinction and can therefore serve as a reliable reference when estimating extinction using other photometric bands.  However, observations of SAGES in the $H\alpha_{\text{n}}$ and $H\alpha_{\text{w}}$ bands have not commenced in the SAGES survey. Consequently, in this work, we rely on alternative data sources to estimate extinction.

We adopted the three-dimensional dust extinction map, Bayestar19 \citep{green20193d}, which covers the sky north of $\text{dec} > -30^{\circ} $ and is therefore well-matched to the SAGES observation area. Stellar distances were estimated using \textit{Gaia} parallaxes, and together with the coordinates, were used to query the Bayestar19 map. 

Following the recommendation of the Bayestar19 documentation\footnote{\url{http://argonaut.skymaps.info/usage}}, we applied Equation \ref{Bayestar19} to convert the Bayestar19 values to the color excess $E(B-V)$,  which is based on conversions involving Pan-STARRS 1 colors and an assumed $R_{v}=3.1$ law \citep{fitzpatrick1999correcting} .

\begin{equation}\label{Bayestar19}
    E(B-V)=0.884\times \text{Bayestar19}.
\end{equation}

Determining the extinction $A_\lambda$ at a specific passband $\lambda$ requires the corresponding extinction coefficient $R_\lambda$. Similarly, to calculate the color excess $E(a-b)$ for a color index $a-b$, the differential extinction coefficient $R_{a-b} = R_a - R_b$ is needed. Extinction coefficients for the $u_{\rm SC}$ and $v_{\rm SAGES}$ bands have been provided by \cite{tan2022stellar}. For the SDSS $g$, $r$, and $i$ bands, extinction coefficients have been derived by \cite{schlafly2011measuring} and \cite{zhang2022empirical}. However, these coefficients are tied to the SFD extinction map \citep{schlegel1998maps}, which has been shown to systematically underestimate $E(B-V)$ values by approximately 14\% \citep{schlafly2011measuring}.

The extinction coefficient for the DDO51 filter has not been previously calculated. One of the traditional methods for calculating the extinction coefficient $R_\lambda$ is the star-pair method(e.g., \citealt{yuan2013empirical}). It assumes that stars with identical atmospheric parameters share the same intrinsic colors and estimates extinction coefficients by comparing the observed colors of star pairs to different extinction levels. In this work, we adopt an approach based on machine learning that is conceptually similar to the star-pair method. Instead of relying on direct star pairings, we train a supervised learning model on a reference sample of low-extinction stars. And we learned a robust mapping from stellar parameters to intrinsic colors. The trained model is then applied to stars in regions of higher extinction to predict their intrinsic colors, leveraging the model's generalization capabilities without requiring exact perfect star pairs.

First, we trained an XGBoost \citep{chen2016xgboost} model to predict intrinsic stellar colors from their atmospheric parameters. Full details of the training details and hyperparameter settings are provided in Appendix B.

The training set for this model consisted of stars selected from our combined APOGEE-\textit{Gaia} XP-SAGES catalog that exhibit low extinction. We identified these low-extinction stars by selecting sources with $E(B-V) < 0.03$, yielding a sample of 19,679 stars. The input features for the model were stellar atmospheric parameters - $T_{\rm eff}$, $\log g$ and [Fe/H], primarily derived from APOGEE high-resolution spectroscopy. These parameter distributions are representative of the overall data set and  adequately cover the parameter range of the higher-extinction stars used in the subsequent extinction coefficient derivation. The target outputs were SAGES color indices: $u_{\rm SC}-v_{\rm SAGES}$, $v_{\rm SAGES}-g$, $g-r$, $r-i$, $g-\text{DDO51}$ and $\text{DDO51}-r$). For these low-extinction stars, the observed colors are assumed to be reliable approximations of their intrinsic colors.

Subsequently, this trained XGBoost model was applied to the remaining sample of stars, which exhibit larger reddening. Using the known atmospheric parameters for each star as input, the model predicted its intrinsic colors - e.g., $(g-i)_0$, $(g-\rm{DDO51} )_0$.
The color excess for each color index $(\lambda_1 - \lambda_2)$ was then computed by subtracting the predicted intrinsic color from the observed color:
\begin{equation}
    E(\lambda_1 - \lambda_2) = (\lambda_1 - \lambda_2)_{\text{obs}} - (\lambda_1 - \lambda_2)_{\text{pred}}
\end{equation}

By performing a linear regression between the calculated color excess $E(\lambda_1 - \lambda_2)$ and the reference $E(B-V)$ for the sample, we derived the differential extinction coefficient $R_{\lambda_1 - \lambda_2} = E(\lambda_1 - \lambda_2) / E(B-V)$ for various SAGES color indices, including those involving the DDO51 filter. These coefficient are commonly reported in the literature, facilitating direct comparisons with previous studies.

However, the extinction coefficient $R_\lambda$ is not a constant. It can vary with stellar parameters and E(B-V) \citep{zhang2022empirical}. To account for potential residual reddening effects and the parameter-dependent nature of extinction not captured by the baseline correction, we included the original $E(B-V)$ estimate as an additional input feature in our final stellar parameter prediction models, alongside the extinction-corrected colors $(\lambda_1 - \lambda_2)_0$. 
This strategy enables the machine learning algorithm to implicitly learn and compensate for these higher-order extinction effects directly from the data.

\subsection{Stellar Surface Gravity Prediction and Uncertainty Assessment}
\subsubsection{Predicting Stellar Surface Gravity Using Machine Learning}

After deriving the extinction coefficients and applying the corresponding  correction to the entire sample, we proceeded to predict the stellar atmospheric parameters. Given the complex, non-linear relationships and interdependencies between intrinsic colors and stellar parameters, we employed machine learning methods for this task. This approach enables us to quantitatively assess the contribution of the DDO51 filter to stellar parameter estimation and to evaluate the performance of the overall SAGES filter system.

We again utilized XGBoost for regression. The input features for the models consisted of the extinction-corrected SAGES color indices: $(g-r)_0$,$(u_{\rm SC}-v_{\rm SAGES})_0$,$(v_{\rm SAGES}-g)_0$,$(r-i)_0$,and $(g-\text{DDO51})_0$, along with the E(B-V) to account for residual reddening and higher-order extinction effects, as discussed above. The data set (LAMOST-\textit{Gaia}-SAGES) was randomly divided into two parts: 80\% was allocated to the training set, which was used exclusively for fitting the parameters of the XGBoost models, and the remaining 20\% was used as the test set for evaluating the model performance.

We trained two XGBoost models: one using extinction-corrected SAGES colors ($u_{\rm SC}$-$v_{\rm SAGES}$), ($v_{\rm SAGES}$-$g$), ($g$-$r$), ($r$-$i$) and E(B-V), and the other incorporating the additional color ($g$-DDO51). Both models used LAMOST DR10 spectroscopic parameters as training labels, allowing us to evaluate and compare the predictive performance under both conditions and further assess our results in the context of the existing literature.

\subsubsection{Evaluating the Impact of Photometric Uncertainty}
To assess the robustness of our log\,$g$ predictions in the presence of observational noise, specifically the impact of photometric uncertainty, we conducted Monte Carlo simulations. This analysis quantifies how the precision of the input photometry, represented by the SNR, propagates to the precision and classification accuracy of the derived log\,$g$.

We simulated various levels of photometric uncertainty by testing magnitude uncertainties ($\sigma_{\text{mag}}$) ranging from 0.01 mag (corresponding to SNR $\approx$ 100) up to 0.2 mag (SNR $\approx$ 5). For simplicity in modeling the impact on color indices, we assumed the magnitude uncertainty was uncorrelated and identical across the two passbands forming a color index ($\lambda_1 - \lambda_2$). Under this assumption, Gaussian noise with a standard deviation of $\sigma_{\text{color}} = \sqrt{2} \times \sigma_{\text{mag}}$ was added to each color index for each star in the independent test set. For each tested $\sigma_{\text{mag}}$, we generated 2,000 perturbed photometric data sets to perform a robust statistical analysis.

For each perturbed data set, the trained XGBoost model was applied to predict log\,$g$ for each star across the 2000 perturbed samples.
From the resulting 2000 log\,$g$ predictions for each star at a given input noise level ($\sigma_{\text{mag}}$), two key metrics were computed:
\begin{enumerate}
    \item Prediction Precision: The standard deviation of the 2000 predicted log\,$g$ values for that star. This quantifies the random error in the log\,$g$ estimate induced by the photometric noise.
    \item Classification Accuracy: For each of the 2000 predictions, we classified the star as a giant (log\,$g < 4.0$) or a dwarf (log\,$g > 4.0$). We then compared this classification to the star's original spectroscopic classification. The accuracy was calculated as the fraction of the 2000 simulations where the photometric classification matched the spectroscopic classification.
\end{enumerate}
The distributions of these metrics across the test set stars as a function of the input magnitude uncertainty ($\sigma_{\text{mag}}$) are presented in Section 4.

\section{Results} \label{sec:Results}

\subsection{Reddening Coefficients}
\begin{figure}
    \centering
    \gridline{
        \fig{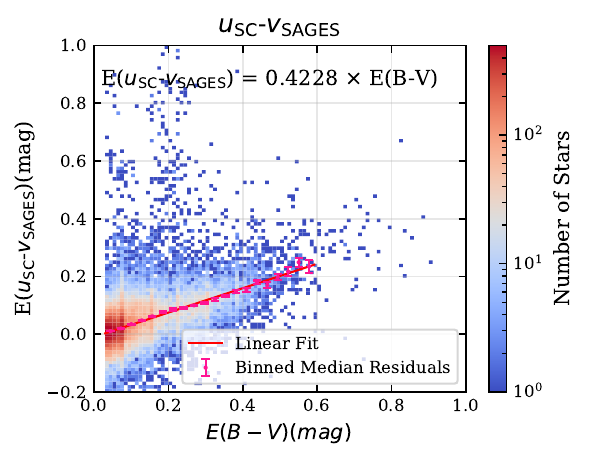}{0.25\textheight}{(a)}
        \fig{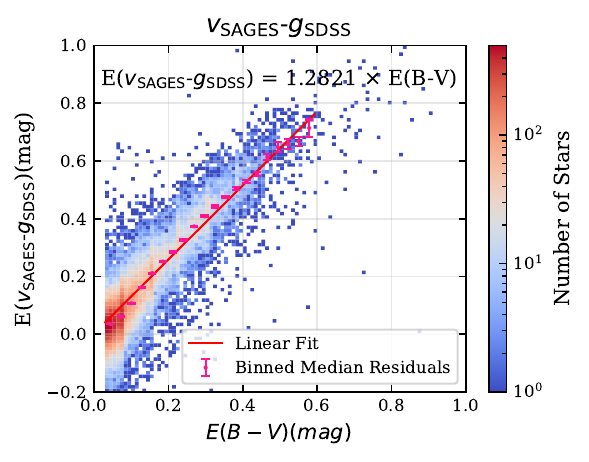}{0.25\textheight}{(b)}
        \fig{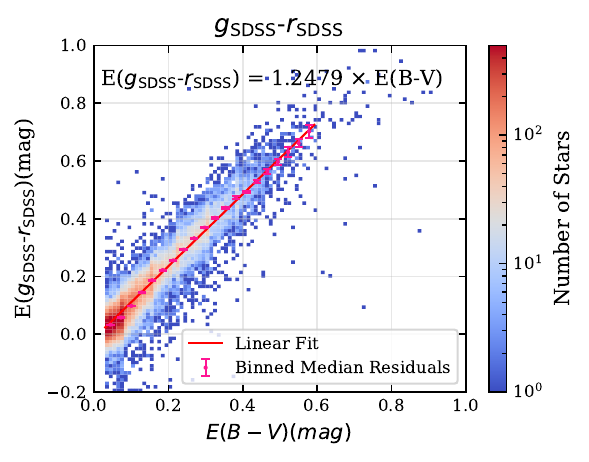}{0.25\textheight}{(c)}
    }
    \gridline{
        \fig{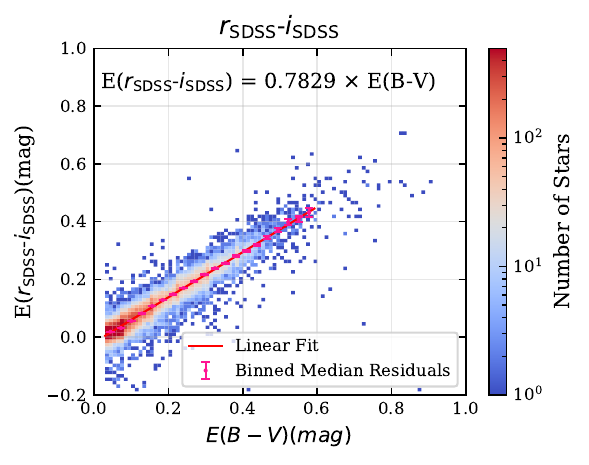}{0.25\textheight}{(d)}
        \fig{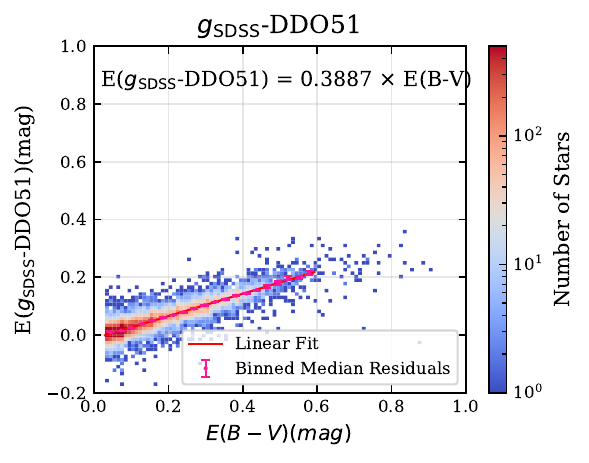}{0.25\textheight}{(e)}
        \fig{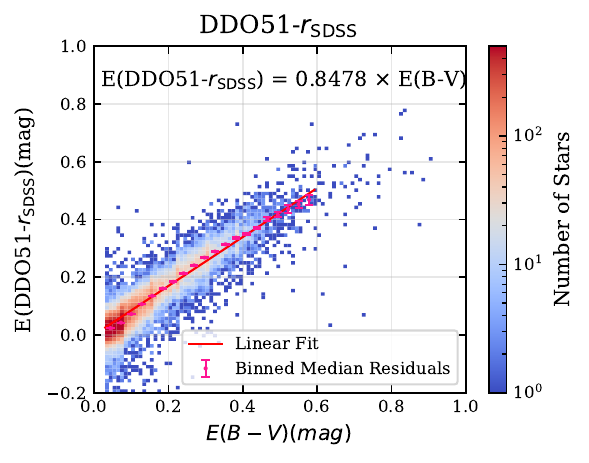}{0.25\textheight}{(f)}
    }
    \caption{Derived color excesses vs. reference $E(B-V)$ from Bayestar19 for various color indices involving SAGES filters. Color excesses were calculated with intrinsic colors predicted by the XGBoost model using APOGEE parameters. The color scale represents the density of stars in the data sample. The red line in each panel shows the best-fit linear relationship $E(\lambda_1 - \lambda_2) = R_{\lambda_1-\lambda_2} \times E(B-V)$, with the derived differential extinction coefficient $R_{\lambda_1-\lambda_2}$ indicated in the legend.}
    \label{fig:ExtinctionCoefficientsXGB}
\end{figure}
\begin{deluxetable*}{lcccc}
\tabletypesize{\scriptsize}
\tablewidth{0pt}
\tablecaption{Comparison of Extinction Coefficients \label{tab:extinction}}
\tablehead{
\colhead{Color} & \colhead{This work} & \colhead{\cite{tan2022stellar}} &
\colhead{\cite{zhang2022empirical}\tablenotemark{a}} & \colhead{\cite{schlafly2011measuring}\tablenotemark{b}}
}
\colnumbers
\startdata
$u_{\text{SC}}-v_{\text{SAGES}}$ & 0.423  & 0.333 & …   & …   \\ 
$g-r$                             & 1.248  & …    & 1.250 & 1.183 \\ 
$r-i$                             & 0.783  & …    & 0.758 & 0.683 \\ 
\enddata
\tablenotetext{a}{Values from the paper's Table 4, select $T_{\rm eff}$=4500 K and $E(B-V)$ =0 }
\tablenotetext{b}{Values are originally relative to $A_{\text{SFD}}$ and have been divided by 0.86 to be consistent with $E(B-V)$ }
\end{deluxetable*}

Figure \ref{fig:ExtinctionCoefficientsXGB} illustrates the derived color excesses for several SAGES color indices, including those incorporating the DDO51 band, plotted against the reference $E(B-V)$ values from Bayestar19. These color excesses were computed using intrinsic colors predicted by the XGBoost model trained on APOGEE atmospheric parameters (see Section \ref{sec:extinction}).
The best-fit linear relationships determined in this work are overplotted on the data. 

For validation purposes, a comparison between our derived extinction coefficients and values previously published is presented in Table \ref{tab:extinction}. The close agreement between our coefficients and those from the literature supports the reliability of our methodology.

\begin{figure}[ht!]
    \centering
    \gridline{
        \fig{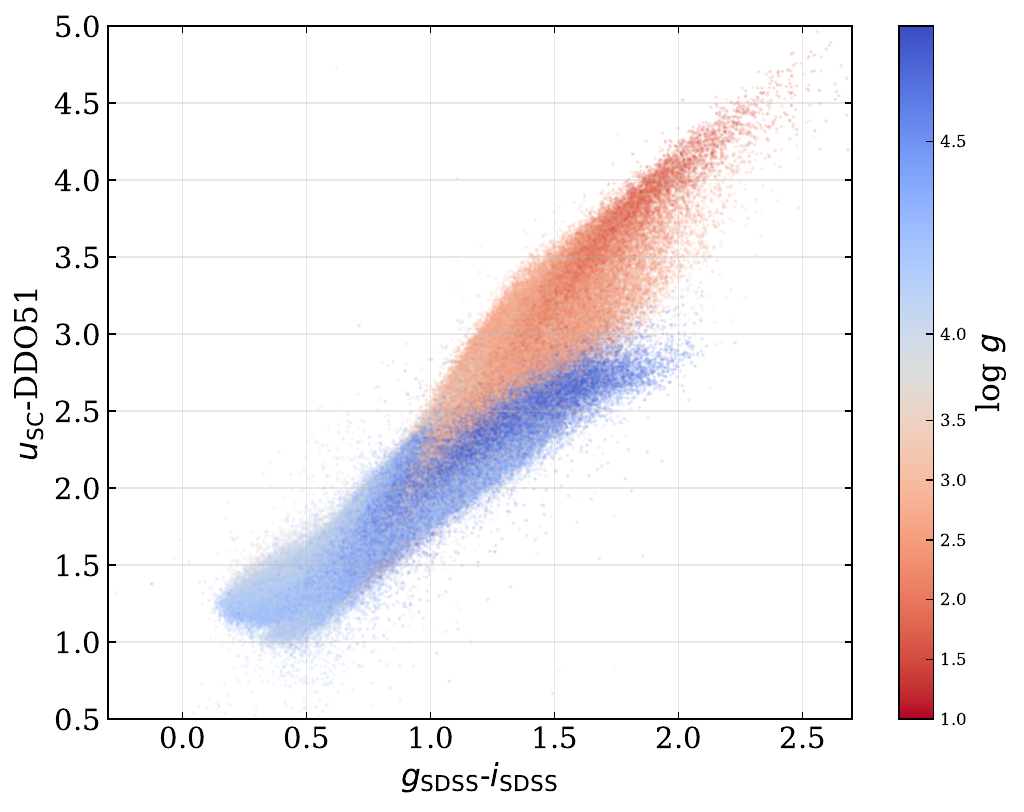}{0.25\textheight}{(a) $u_{\rm SC}$-DDO51}
        \fig{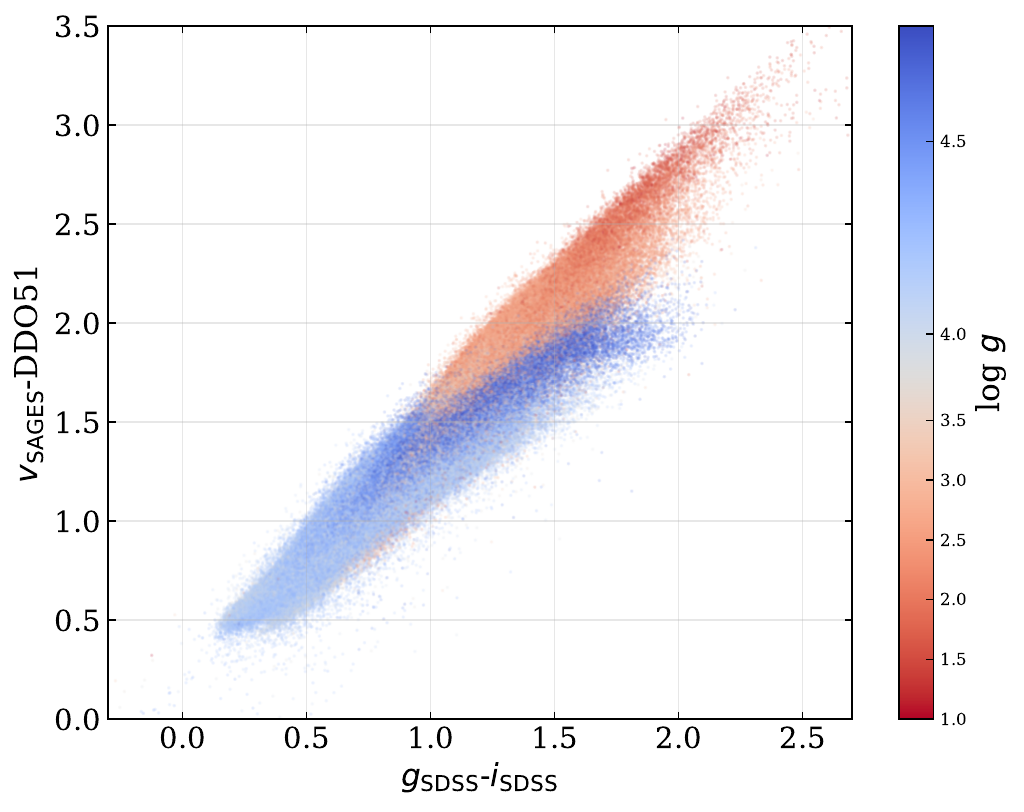}{0.25\textheight}{(b) $v_{\rm SAGES}$-DDO51}
        \fig{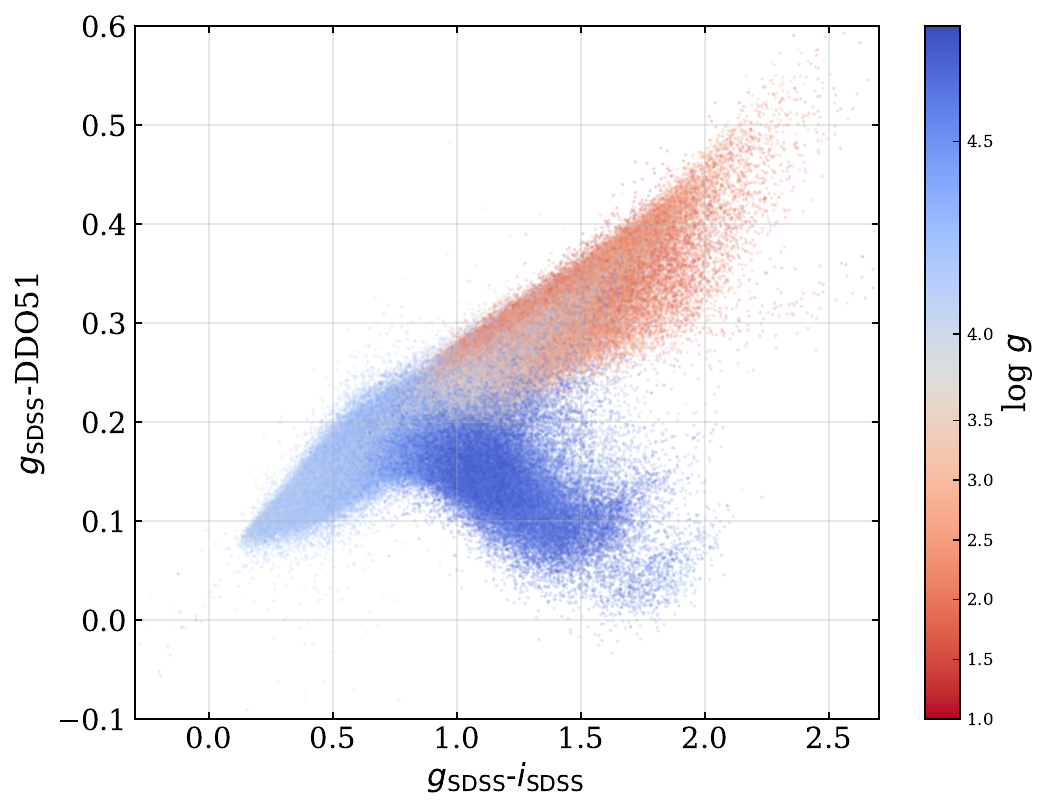}{0.25\textheight}{(c) $g$-DDO51}
    }
    \caption{Color-color diagrams constructed from observed magnitudes prior to extinction correction. Each panel shows a DDO51-related color index plotted against the $g-i$ color: (a) $u_{sc}$-DDO51, (b) $v_{\rm SAGES}$-DDO51, (c)$g$-DDO51. The points are color-coded by log\,$g$, with values taken from the LAMOST DR10 catalog, as indicated by the color bar.
}
    \label{fig:colormap1}
\end{figure}

\begin{figure}[ht!]
    \centering
    \gridline{
        \fig{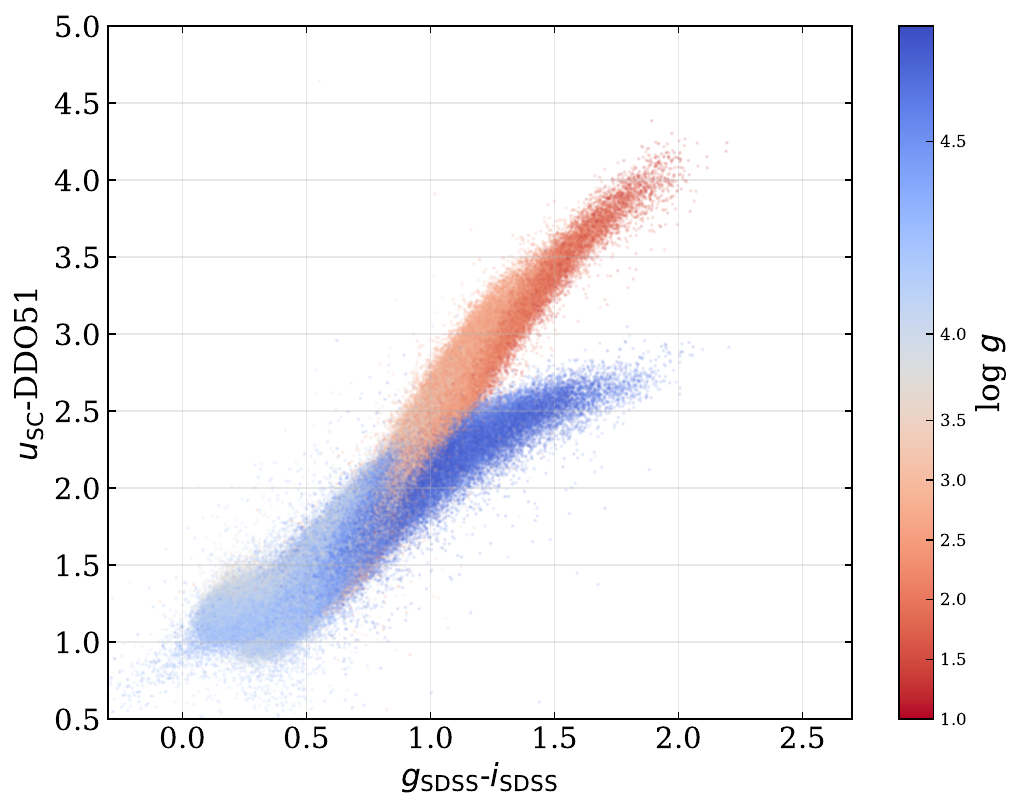}{0.3\textwidth}{(d) $u_{\rm SC}$-DDO51}
        \fig{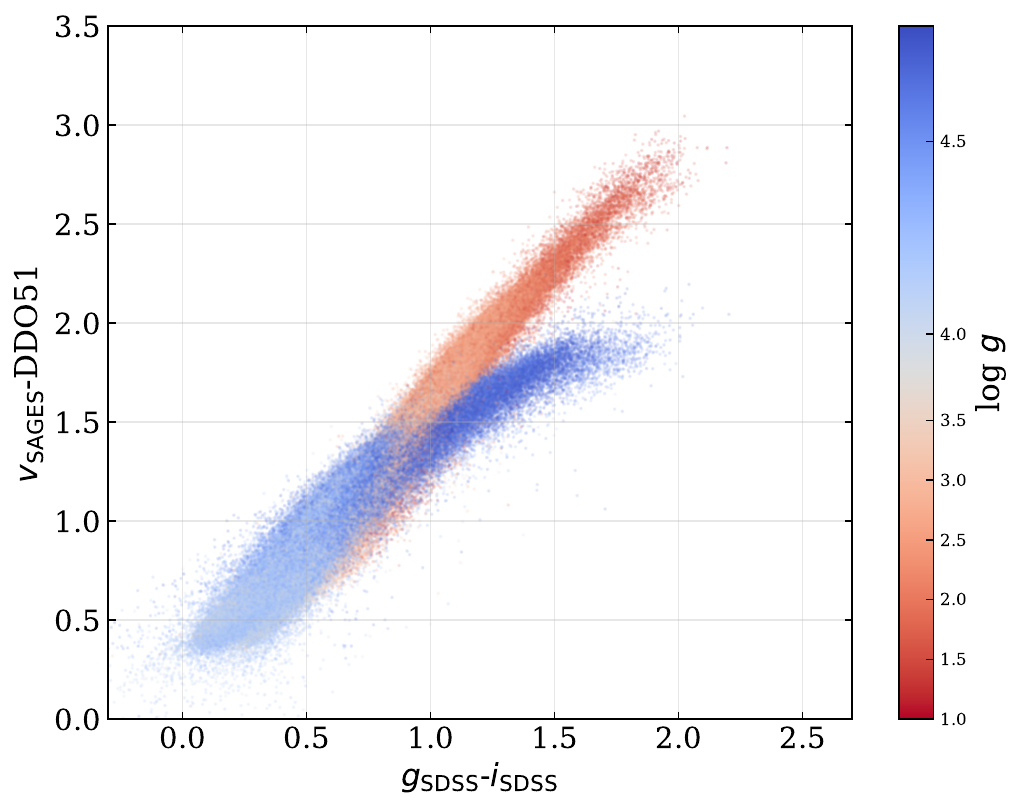}{0.3\textwidth}{(e) $v_{\rm SAGES}$-DDO51}
        \fig{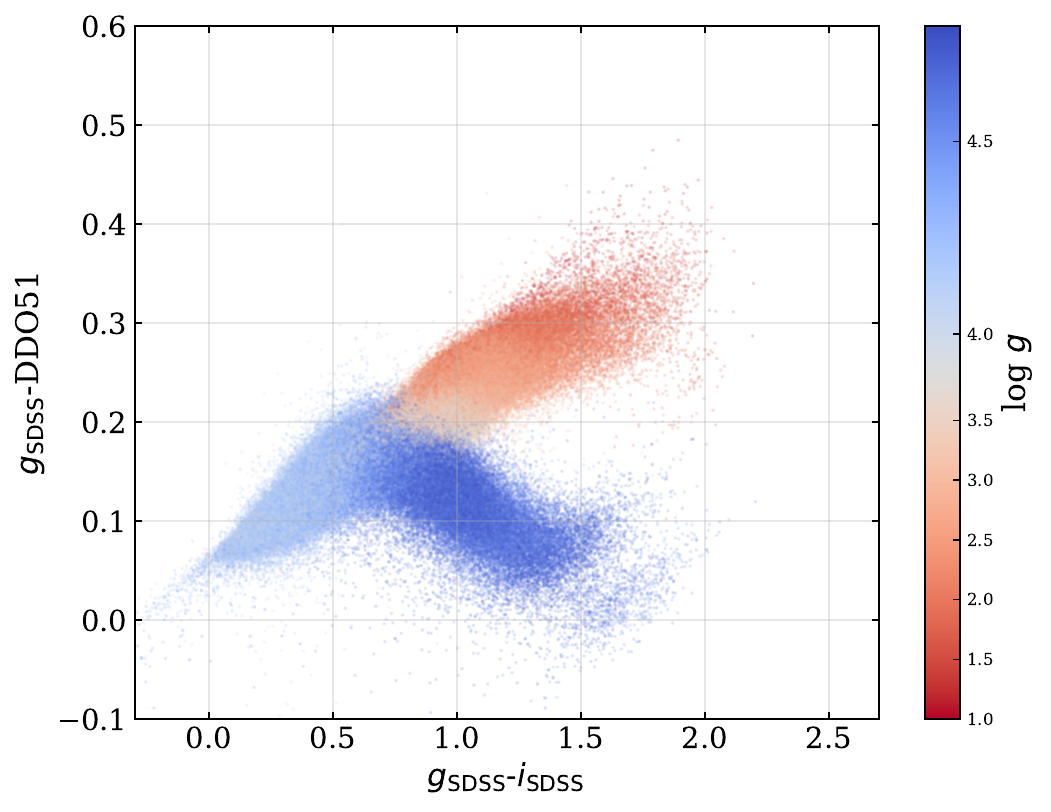}{0.3\textwidth}{(f) $g$-DDO51}
    }
    \caption{Same as Figure \ref{fig:colormap1}, but after applying the extinction correction.
}
    \label{fig:colormap2}
\end{figure}

A comparison between the observed (Figure \ref{fig:colormap1}) and extinction-corrected (Figure \ref{fig:colormap2}) color-color diagrams highlights the significant impact of interstellar reddening and the effectiveness of our correction method. 
The considerable scatter caused by extinction present in the Figure \ref{fig:colormap1}, which disperses stars with similar log\,$g$, is markedly reduced after correction. This results in  more tightly clustered and well-defined stellar sequences in Figure \ref{fig:colormap2}, demonstrating that the applied correction effectively  mitigates reddening-induced displacements. For $( g-i)_0 >$ 0.8, a clear separation between giants and dwarfs is observed, demonstrating the DDO51 filter's potential to classify late-type stars based on surface gravity. Furthermore, comparing these correction colors with theoretical colors from model atmospheres (Appendix A) reveals a qualitative agreement in the overall trends and the separation of different log\,$g$ populations. However, noticeable quantitative discrepancies remain between the observed stellar loci and theoretical predictions, indicating potential limitations in the synthetic spectra, residual uncertainties in observational data or extinction corrections, or complexities in stellar atmospheres that are not fully captured.

\subsection{Regression Results}

\subsubsection{Surface Gravity Predictions with and without the DDO51 Filter}
\begin{figure}[ht!]
\gridline{
\fig{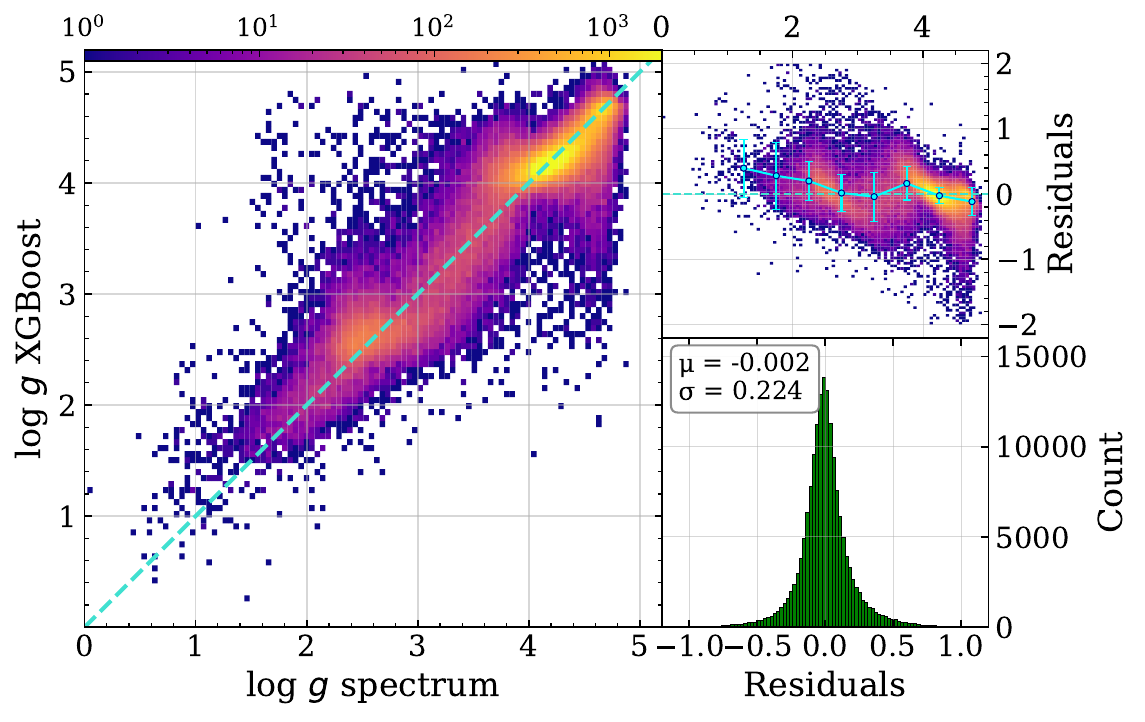}{0.49 \textwidth}{(a) without DDO51 filter}
\fig{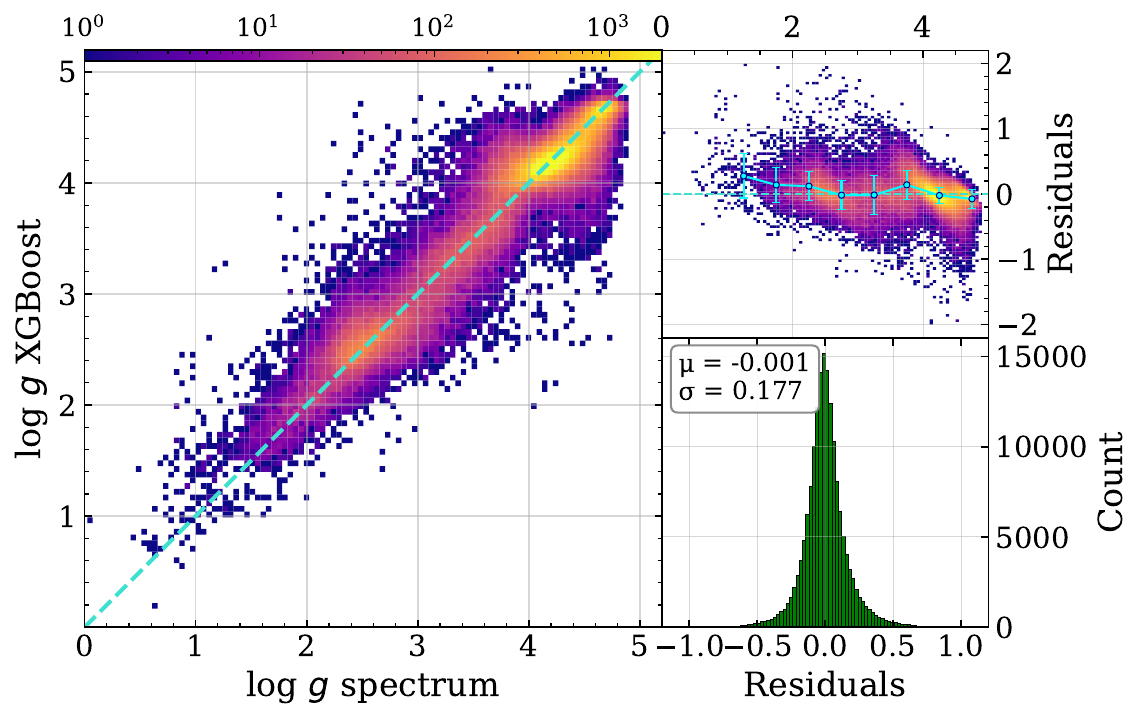}{0.49\textwidth}{(b) with DDO51 filter}
         }
\caption{Comparison of log\,$g$ predictions from XGBoost models  trained on SAGES and \textit{Gaia} XP synthetic photometry, with and without the DDO51 filter. (a): Results using SAGES colors and E(B-V), without the DDO51 filter. (b): Results with the same features as panel (a), but additionally including the ($g$-DDO51) color. \textit{Large panels}: 2D histograms of XGBoost predicted vs. spectroscopic log\,$g$ from LAMOST, The color scale indicates stellar density (logarithmic scale). The cyan dashed line indicates the one-to-one relation. \textit{Small top panels}: Standard deviation of residuals (predicted – spectroscopic) log\,$g$  as a function of spectroscopic log\,$g$, binned in 0.5 dex intervals. The cyan line and error bars indicate the median and standard deviation within each bin. \textit{Small bottom panels}: Histograms of residuals, annotated with mean ($\mu$) and standard deviation ($\sigma$).
\label{fig:logg_comparison}}
\end{figure}

Figures \ref{fig:logg_comparison}(a) and (b) show the performance of the two models in predicting log\,$g$, respectively. A clear improvement is observed when the DDO51 filter is included, as evidenced by the tighter distribution around the one-to-one reference line in dashed. In contrast, the model without the DDO51 filter (Fig. \ref{fig:logg_comparison}a) shows a larger scatter and a systematic tendency to underestimate log\,$g$ for a subset of dwarf stars. This underestimation leads to increased misclassification of dwarfs as giants. The inclusion of the DDO51 filter (Fig. \ref{fig:logg_comparison}b) significantly mitigates this issue, reducing both scatter and classification error.

The overall standard deviation of the residuals (predicted log\,$g$ $-$ LAMOST log\,$g$) decreases from 0.224 dex without the DDO51 filter to 0.177 dex with it, representing a 21.0\% improvement. Both models exhibit a slight tendency to underestimate log\,$g$ for dwarf stars (log\,$g$ \textgreater\ 4) and overestimate it for giant stars. but this systematic trend becomes less pronounced when including the DDO51 filter, as detailed below (Figure \ref{fig:logg_comparison2}).

\subsubsection{Dependence on Stellar Parameters Performance}
\begin{figure}[ht!]
\gridline{\fig{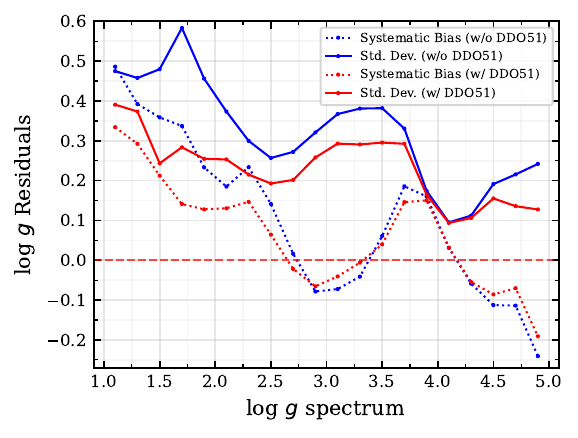}{0.32\textwidth}{(a) log\,$g$ Residuals vs. log\,$g$ spectrum}
          \fig{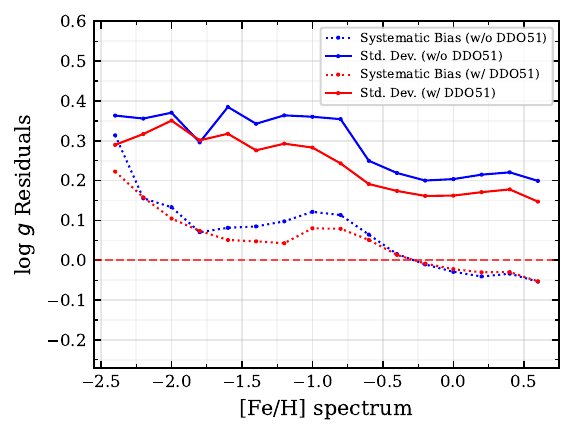}{0.32\textwidth}{(b) log\,$g$ Residuals vs. [Fe/H]}
          \fig{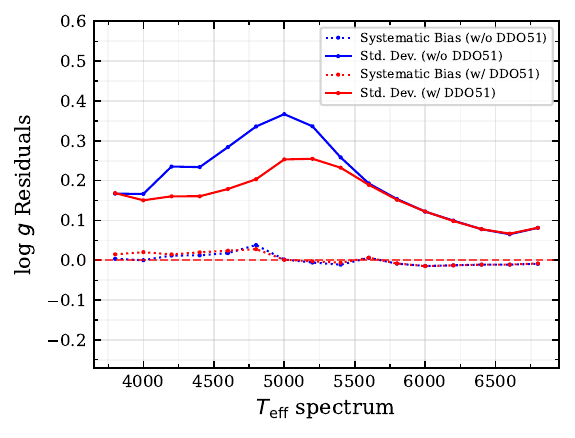}{0.32\textwidth}{(c) log\,$g$ Residuals vs. $T_{\rm eff}$}}
\caption{Residual analysis of log\,$g$ predictions as a function of spectroscopic parameters, with and without the DDO51 Filter. (a) Residuals (predicted log\,$g$ - spectroscopic log\,$g$) plotted against spectroscopic log\,$g$. (b) Residuals vs. [Fe/H]. (c)  Residuals vs. against $T_{\rm eff}$. In all panels, solid lines show the standard deviation of the residuals, and dotted lines indicate the mean of residual within bins. Bin widths are 0.2 dex for log\,$g$, 0.2 dex for [Fe/H], and 200 K for $T_{\rm eff}$. Blue lines represent the model without DDO51, while red lines represent the model including DDO51.
\label{fig:logg_comparison2}}
\end{figure}

Figure \ref{fig:logg_comparison2} provides a detailed residual analysis, examining the behavior of the residuals as a function of spectroscopic log\,$g$ (a), [Fe/H] (b), and $T_{\rm eff}$ (c). Across nearly the entire parameter space, the inclusion of the DDO51 filter leads to a reduction in both the standard deviation of the residuals (solid lines) and the systematic bias (dotted lines), demonstrating enhanced precision and reduced systematic error in the log\,$g$ estimates. 

The most pronounced improvements are observed for giant stars with log $g <$ 2.5 and dwarf stars with log $g >$ 4.5.										
While both models exhibit systematic trends across the log\,$g$ range, these biases are notably diminished when DDO51 photometry is included, especially at the extremes.
Specifically, Figure \ref{fig:logg_comparison2}(b) shows that the DDO51 filter improves log\,$g$ predictions for stars with [Fe/H] $>$ -2, with particularly noticeable enhances for metal-poor stars in the metallicity range of -2 $<$ [Fe/H] $<$ -1. Figure \ref{fig:logg_comparison2}(c) reveals that the DDO51 filter significantly improves log\,$g$ predictions for stars with $T_{\rm eff} < 5600$ K, while having a negligible effect on stars hotter than 5600 K. This behavior aligns with theoretical expectations, as the \ion{Mg}{1} \,b triplet, the primary spectral feature targeted by the DDO51 filter, is theoretically predicted to be stronger in the atmospheres of cooler stars.

\begin{figure}[ht!]
\gridline{
\fig{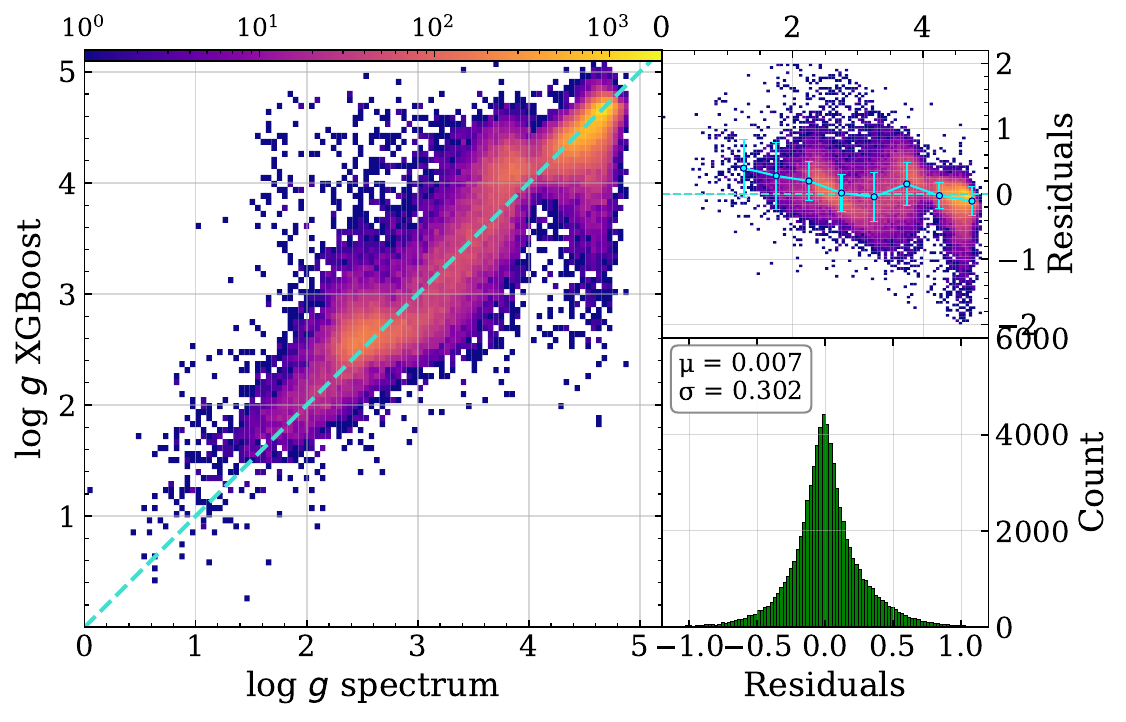}{0.49 \textwidth}{(a) without DDO51 filter}
\fig{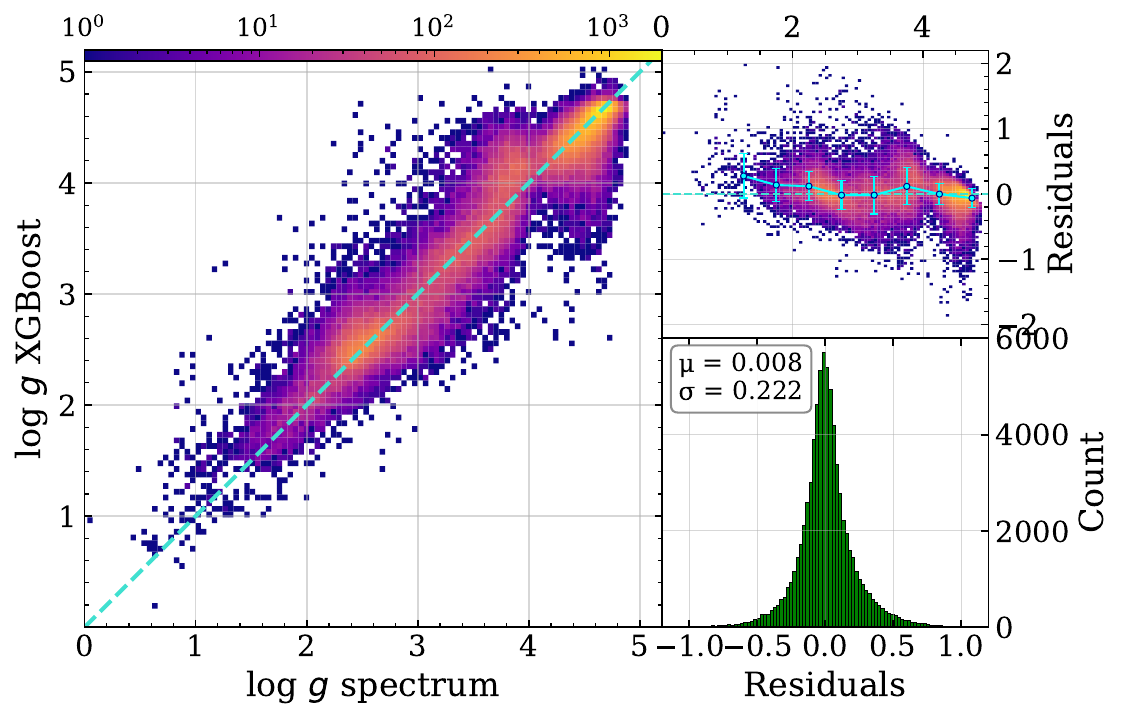}{0.49\textwidth}{(b) with DDO51 filter}
         }
\caption{Performance of the XGBoost models from Figure \ref{fig:logg_comparison} for log\,$g$ prediction, restricted to GK-type stars with effective temperatures below 5600 K. (a) Results using SAGES colors and E(B-V) only. (b) Results including the ($g$-DDO51) color. The layout and symbols are the same as in Figure \ref{fig:logg_comparison}.
\label{fig:logg_comparison3}}
\end{figure}

To isolate the filter's impact on these cooler stars, Figure \ref{fig:logg_comparison3} replicates the comparison from Figure \ref{fig:logg_comparison}, but focuses exclusively on GK-type stars ($T_{\rm eff} < 5600 \rm K$ ). The improvement from including DDO51 is visually pronounced. The standard deviation of the overall residuals for this cool subsample decreases from 0.302 dex to 0.222 dex, corresponding to a 26.5\% improvement – slightly exceeding the 21.0\% improvement observed in the full sample.

\subsubsection{Performance and Limitations for Metal-Poor Stars}

\begin{figure}[ht!]
\gridline{
\fig{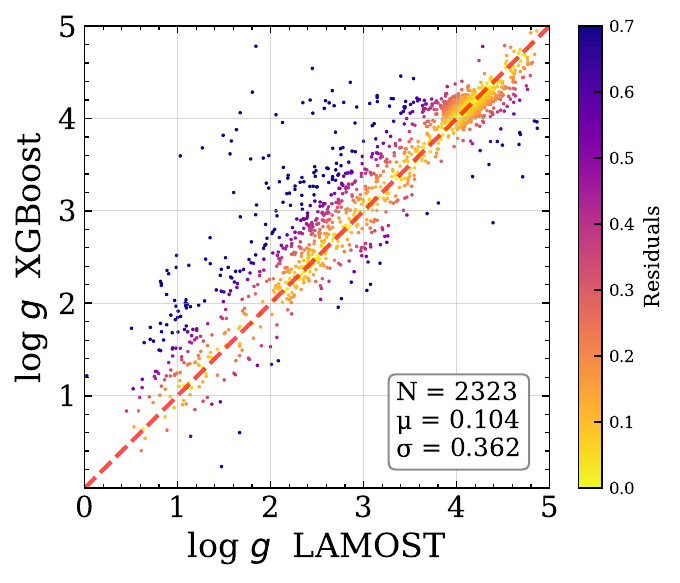}{0.49 \textwidth}{(a) without DDO51 filter}
\fig{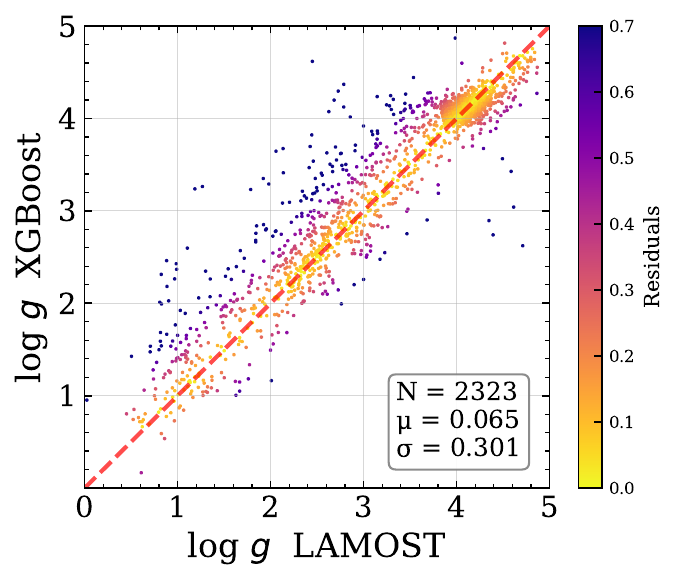}{0.49\textwidth}{(b) with DDO51 filter}
         }
\caption{Performance of the XGBoost models  on a subsample of 2,323 metal-poor stars ([Fe/H] $<$ -1.0) from the test set. (Left): Predictions without the DDO51 filter. (Right): Predictions including the DDO51. The color of each point represents the absolute residual between the predicted and spectroscopic values.
}
\label{fig:logg_mp_comparison}
\end{figure}

\begin{figure}[ht!]
\gridline{
\fig{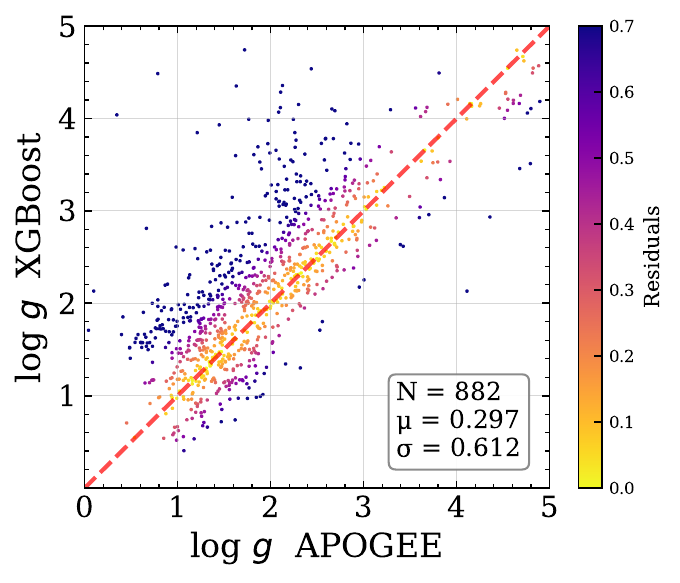}{0.49 \textwidth}{(a) without DDO51 filter}
\fig{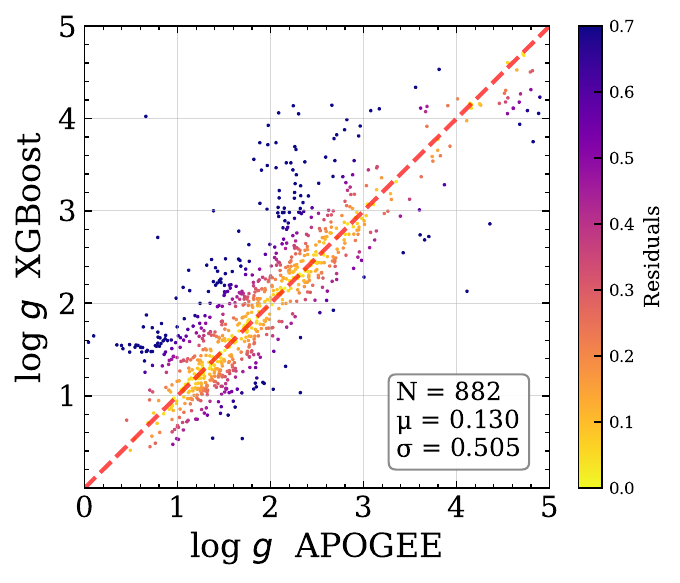}{0.49\textwidth}{(b) with DDO51 filter}
         }
\caption{Cross-survey validation of the model's generalization capability on an independent sample of 882 metal-poor ([Fe/H] $<$ -1.0) stars from the APOGEE survey. The panels compare log g predicted by our LAMOST-trained model against APOGEE's spectroscopic values. The
layout and symbols are the same as in Figure \ref{fig:logg_mp_comparison}.}

\label{fig:logg_mp_comparison_apogee}
\end{figure}

Due to the intrinsic rarity of metal-poor stars ([Fe/H] $< -1.0$), they are sparsely represented in our dataset, as illustrated in Figure \ref{fig:catalogs}. To evaluate the effectiveness of the DDO51 filter for metal-poor stars, we performed a dedicated analysis on a selected subsample.

We first evaluated the model performance on the metal-poor stars in our test set. As shown in Figure \ref{fig:logg_mp_comparison}, we identified 2,323 stars with LAMOST spectroscopic metallicities of [Fe/H] $<$ -1.0.
Without the DDO51 filter, the residuals exhibit a standard deviation of 0.362 dex, and a systematic bias of 0.104 dex. When the DDO51 filter is included, the precision improves to a standard deviation of 0.301 dex (16.8\%) and the bias is reduced to μ = 0.065 dex (37.5\%).

It is noteworthy that the absolute accuracy of log $g$ prediction for metal-poor stars remains lower than that for the overall sample, primarily due to their sparse representation in the training data and the larger uncertainties in the LAMOST labels. Nevertheless, this does not alter our conclusion that incorporating the DDO51 filter enhances the precision for metal-poor stars.

To evaluate the generalization capability of our model and confirm that it learns robust physical features rather than overfitting to systematics in the LAMOST training data, we tested its performance on an external dataset from APOGEE. We selected stars from the APOGEE catalog that were not used in the training set and identified 882 stars with APOGEE [Fe/H] $<$ -1.0. We then applied our LAMOST-trained models to this external sample.

Figure \ref{fig:logg_mp_comparison_apogee} shows that the model predictions against the APOGEE values. The results confirm the trend observed in the LAMOST test set. Without DDO51, the comparison yields a large standard deviation ($\sigma$ = 0.612 dex) and a substantial bias ($\mu$ = 0.297 dex). Including DDO51 improves the results, reducing the standard deviation by 17.5\% to 0.505 dex and especially reducing the bias by 56.2\% to 0.130 dex.

While the absolute discrepancies remain large ($\sigma \approx$ 0.5 dex), this reflects the combined effects of the inherent systematic differences between the LAMOST and APOGEE spectroscopic scales (as discussed in Section \ref{external_vail} and Figure \ref{fig:logg_external}) and the challenges of extrapolation into a sparsely sampled parameter space.  Therefore, the key takeaway is not the absolute error, but the relative improvement of 17.5\% in precision and 56.2\% in bias, which powerfully demonstrates the DDO51 filter's value even in a challenging cross-survey context.

In summary, despite the limitations imposed by the scarcity of metal-poor stars in the training data, the inclusion of the DDO51 filter consistently improves the estimation of log g for stars with [Fe/H] $<$ -1.0. This validates the physical basis of the DDO51 filter—its sensitivity to the Mg I b triplet—and confirms its utility in mitigating biases for metal-poor stars.

\subsubsection{External Validation and Impact of Photometric Uncertainty on Predictions}\label{external_vail}
\begin{figure}[ht!]
\gridline{
\fig{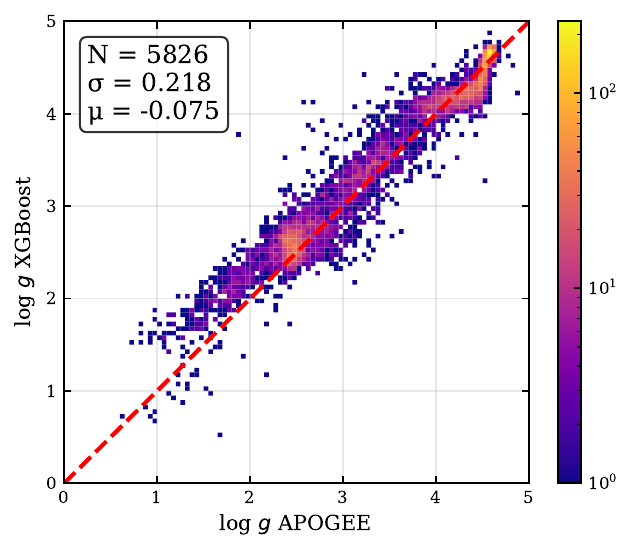}{0.49 \textwidth}{(a) XGBoost vs APOGEE}
\fig{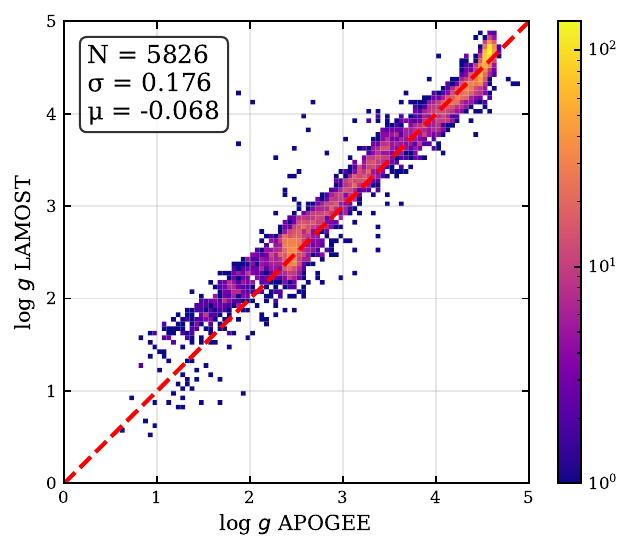}{0.49\textwidth}{(b) LAMOST vs APOGEE}
         }
\caption{Evaluation of XGBoost log\,$g$ predictions and comparison with APOGEE differences, using stars common to our test set and APOGEE DR17. (Left): XGBoost-predicted log\,$g$ from our model (including DDO51) plotted against APOGEE DR17 spectroscopic log\,$g$. (Right): LAMOST DR10 spectroscopic log\,$g$ plotted against APOGEE DR17 spectroscopic log\,$g$. This panel illustrates the inherent systematic differences between the LAMOST and APOGEE spectroscopic determinations for the same set of stars. The red dashed line in both panels represents the one-to-one relation. The standard deviation ($\sigma$) and systematic bias ($\mu$) of the differences are shown.
\label{fig:logg_external}}
\end{figure}

To provide external validation, we compared the predicted log\,$g$ values from out XGBoost model (including the DDO51 filter) to independent spectroscopic measurements from APOGEE DR17 for 5,826 stars common to our test set. Figure \ref{fig:logg_external}(a) shows this comparison. The standard deviation of the differences (XGBoost prediction - APOGEE) is 0.218 dex, with a systematic bias (mean offset) of -0.075 dex.

It is crucial to acknowledge that part of this discrepancy arises from systematic differences between the LAMOST and APOGEE spectroscopic surveys. Figure \ref{fig:logg_external}(b) illustrates these inherent differences by directly comparing the LAMOST DR10 and APOGEE DR17 spectroscopic log\,$g$ values for the same sample of stars. The resulting standard deviation is 0.176 dex, with a systematic bias of -0.068 dex (LAMOST - APOGEE). Assuming the LAMOST and APOGEE differences represent the baseline systematic uncertainty between spectroscopic surveys, the additional scatter introduced by our photometric prediction method can be estimated as $\sqrt{0.218^2 - 0.176^2} \approx 0.129$ dex. This suggests that the XGBoost model contributes an additional random uncertainty of approximately 0.13 dex in log\,$g$ beyond the inter-survey systematics.

\begin{figure}[ht!]
\gridline{
\fig{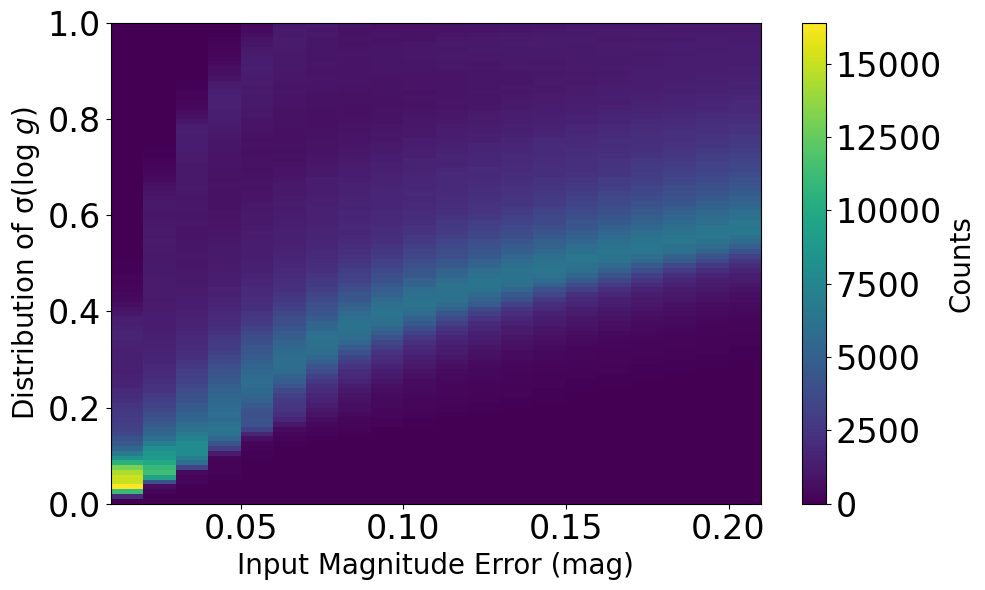}{0.49 \textwidth}{(a) Standard deviation of predicted log\,$g$ vs. magnitude uncertainty.}
\fig{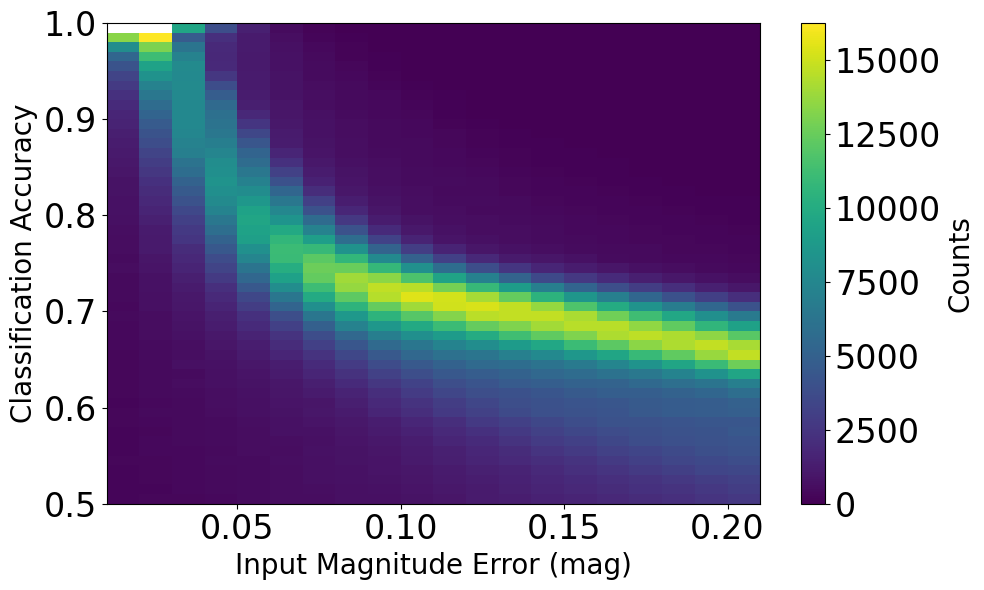}{0.49\textwidth}{(b): Classification accuracy vs. magnitude uncertainty.}
         }
\caption{Impact of input magnitude uncertainty on the precision and accuracy of stellar parameter estimation, assessed via Monte Carlo simulations. (a) Distribution of the standard deviation of predicted log\,$g$ values as a function of input magnitude uncertainty. Gaussian noise with varying standard deviations (x-axis) was added to the input magnitudes, and the standard deviation of the resulting log\,$g$ predictions (from 2,000 simulations) is shown for each star. (b) Distribution of giant/dwarf classification accuracy as a function of input magnitude uncertainty. The fraction of simulations in which the predicted log\,$g$ classification (giant if log $g <$ 4.0, dwarf if log $g >$ 4.0) agrees with the spectroscopic classification is shown. The 2D histograms color scale in both panels represents the number of stars within each bin.
\label{fig:mc}}
\end{figure}

Figure \ref{fig:mc} presents the results of Monte Carlo simulations, with panel (a) showing that the distribution of the standard deviation in predicted log\,$g$ values, and panel (b) displaying the distribution of classification accuracies. As expected, a positive correlation is observed between photometric uncertainty and the standard deviation of the predicted log\,$g$. Specifically, increasing the input magnitude uncertainty from 0.01 mag to 0.1 mag results in a corresponding rise in the log\,$g$ prediction standard deviation - from typically less than 0.1 dex to approximately 0.5 dex. At the highest tested uncertainty of 0.2 mag, the standard deviation reaches around 0.6 dex. Concomitantly, classification accuracy declines with increasing photometric noise. Accuracy drops from nearly 100\% at 0.01 mag uncertainty to approximately 80\% at 0.05 mag, and further down to about 70\% at 0.1 mag uncertainty. These results highlight the critical importance of high-precision photometry for reliable surface gravity estimation.

\section{Discussion}
Our results demonstrate that incorporating the DDO51 filter alongside the other SAGES passbands significantly enhances the accuracy of surface gravity estimation. This finding provides strong quantitative support for the physical rationale behind the DDO51 filter's design, as discussed in Section~\ref{sec:intro} and illustrated in Figure~\ref{fig:DDO51}.

\subsection{Physical Basis for DDO51 Sensitivity}

The observed improvements in surface gravity estimation directly validate the effectiveness of targeting the pressure-broadened \ion{Mg}{1} \,b triplet absorption bands, both of which are highly sensitive to surface gravity in G- and K-type stars. The DDO51 filter, centered on these spectral features, captures critical information that is largely orthogonal to that provided by broad-band color indices. For instance, a cool, metal-poor giant and a hotter, metal-rich dwarf can exhibit similar broad-band colors (e.g., in $(g - r)$ or $(r - i)$), yet can be clearly distinguished using DDO51-based indices due to their differing line-broadening characteristics.

\subsection{Comparison with Other Methods}

The $\log g$ precision achieved in this work—0.180 dex relative to LAMOST—is competitive with, and in many cases superior to, several existing photometric methods. Classical Strömgren photometry, for example, utilizes the $c_1$ index, defined as $(u - v) - (v - b)$, which is primarily sensitive to the Balmer jump. While this method is effective for hotter A- and F-type stars, it becomes significantly less reliable for cooler G- and K-type stars due to the diminishing strength of the Balmer discontinuity at lower effective temperatures \citep{stromgren1966spectral}. Similarly, the SAGES $u_{\rm SC}$ and $v_{\rm SAGES}$ bands are subject to similar limitations, as they probe the same underlying physical regime. In contrast, the DDO51 filter provides a robust gravity diagnostic for cooler stars by probing pressure-broadened lines that remain prominent at lower effective temperatures.

An alternative approach for estimating surface gravity combines multi-band photometry with precise \textit{Gaia} parallaxes to estimate absolute magnitudes, from which log\,$g$ can be inferred. Under favorable conditions, this method can achieve precisions of approximately 0.2--0.3 dex \citep[e.g.,][]{lin2022distances, anders2022photo, zhang2025photometric}. However, its reliability is strongly dependent on low fractional parallax uncertainties, typically less than $10-20\%$. For more distant stars, the growing astrometric errors degrade the accuracy of this method substantially.

Compared to other narrow- or medium-band photometric surveys such as J-PLUS \citep[$\sim$0.15 dex;][]{yang2022j} or those leveraging full-resolution Gaia XP spectra \citep[e.g.,][]{khalatyan2024transferring, andrae2023Gaia, ye2025mapping, fallows2024stellar}, our method offers a significant advantage in both data dimensionality and operational efficiency. While XP-based approaches utilize 1,536 flux bins per star, our model requires only six photometric features — two directly observed SAGES colors and four synthesized color indices. Despite this simplicity, our method achieves comparable performance, underscoring the diagnostic strength of strategically designed filters such as DDO51 in enabling efficient, low-dimensional inference of stellar parameters.

\subsection{Limitations and Caveats}

Several factors should be taken into account when interpreting the results of this study.

Firstly, this work relies on synthetic DDO51 photometry derived from \textit{Gaia} XP spectra. Due to the limited spectral resolution (R $\lesssim$ 50) and intrinsic measurement uncertainties in the XP data, the accuracy of the synthesized magnitudes is fundamentally constrained. Moreover, the synthesis process does not account for the precise instrumental response function or atmospheric transmission profile specific to the DDO51 filter. As a result, the synthetic magnitudes may deviate systematically from those obtained through direct observations, introducing uncertainties that are absent in directly measured narrow-band photometry.

Secondly, the forthcoming SAGES DDO51 observations will face practical implementation challenges. These include field-dependent zero-point calibration, spatially variable atmospheric extinction, and minor but non-negligible variations in filter transmission profiles. Such instrumental and environmental effects — absent in our synthetic framework — are expected to degrade both the photometric precision and classification accuracy. Consequently, the performance reported in this work should be interpreted as an optimistic upper limit, representing best-case expectations rather than guaranteed outcomes for the final SAGES DDO51 catalog.

Thirdly, our machine learning models are inherently data-driven. Their predictive performance is fundamentally limited by the quality, consistency, and parameter-space coverage of the training data sets (LAMOST and APOGEE). Systematic errors in the spectroscopic labels propagate directly into the inferred photometric parameters. Furthermore, the models may generalize poorly to stellar populations that lie outside the distribution of the training data—such as extremely metal-poor stars, ultra-cool dwarfs, or chemically peculiar objects. The use of separate spectroscopic data sets to train distinct components of the model (e.g., LAMOST for $\log g$, APOGEE for extinction), though pragmatically motivated by data availability, could also introduce subtle inconsistencies across the inferred photometric parameter scales.

Finally, we identified an additional random uncertainty of approximately 0.13 dex  in $\log g$ when comparing our predictions to APOGEE, which exceeds the typical LAMOST–APOGEE discrepancy (Figure~\ref{fig:logg_external}). This excess scatter likely arises from several contributing factors: (i) residual inaccuracies in the synthetic photometry pipeline; (ii) limitations in the machine learning model's capacity to fully capture complex nonlinear relationships between color indices and stellar parameters; and (iii) unmodeled stellar physics, such as unresolved binarity, stellar rotation, magnetic activity, or variations in $\alpha$-element abundances. In addition, systematic offsets between the photometric and spectroscopic parameter scales—beyond the primary atmospheric parameters ($T_{\rm eff}$, log\,$g$, [Fe/H])—may also contribute.

Despite these limitations, it is worth noting that the expected photometric depth of the SAGES survey (Table~\ref{tab:SAGES_limits}) extends significantly fainter than the magnitude limits of the \textit{Gaia} XP spectra. As demonstrated by our Monte Carlo simulations (Figure~\ref{fig:mc}), for stars brighter than $G \sim 17.5$ mag, surface gravity predictions maintain a precision better than 0.2 dex, with giant/dwarf classification accuracies exceeding 85\%. This indicates that, within this brightness regime, SAGES narrow-band photometry has strong potential to yield robust stellar parameter estimates, even in the absence of high-resolution spectroscopy or precise astrometry.

\section{Conclusion}

This study assessed the contribution of the DDO51 intermediate-band filter to stellar surface gravity estimation within the SAGES photometric system. We combined observed SAGES ($u_{\rm SC}, v_{\rm SAGES}$) photometry with synthetic \textit{Gaia} XP data ($g, r, i$, DDO51) applying machine learning techniques to perform extinction correction (including deriving the DDO51 coefficient) and to train XGBoost models on spectroscopic labels LAMOST and APOGEE. This framework enabled a quantitative assessment of the DDO51 filter's impact on surface gravity predictions.

Our findings demonstrate that incorporating DDO51 significantly improves log\,$g$ estimation, reducing the residual standard deviation relative to LAMOST spectroscopy by 21.0\% to 0.177 dex overall (and by 26.5\% for GK-type stars, $3750\rm K <T_{\rm eff} < 5600 \rm K$ ). This improves the distinction between giants and dwarfs, reducing misclassification rates, as illustrated in Figure~\ref{fig:logg_comparison}. Although the DDO51 filter targets a metallic feature (the Mg I b triplet), the effectiveness is validated even for metal-poor stars ([Fe/H]$<$-1.0). The improvement in log g precision is in accordance with the entire sample, and proving particularly valuable for mitigating systematic bias, which was reduced by 37.5\% (internal test set) and 56.2\% (external validation set). 

Compared to traditional photometric indices with \textit{Gaia} parallax and even high-dimensional XP-based approaches, our method achieves highly competitive precision by leveraging only six carefully selected photometric bands, including DDO51. This underscores the diagnostic power of well-designed intermediate-band filters and demonstrates the feasibility of efficient, scalable ground-based stellar parameter estimation with significant precision.

As SAGES begins acquiring real DDO51 observations, our current results provide an optimistic performance benchmark. Future work will involve retraining models using real photometric data and extending the framework to fainter stars and rarer stellar populations, leveraging the full depth and breadth of the SAGES survey.

\section{Acknowledgements}
This work is supported by the National Natural Science Foundation of China (NSFC) under grant No.12588202, National Key R\&D Program of China No.2023YFE0107800, No.2024YFA1611900

ZG acknowledges funding under Department of Human Resources and Social Security of Xinjiang Uygur Autonomous Region Introduced Project “Tianchi talent”.

This work has made use of data from the European Space Agency (ESA) mission
{\it Gaia} (\url{https://www.cosmos.esa.int/Gaia}), processed by the {\it Gaia}
Data Processing and Analysis Consortium (DPAC,
\url{https://www.cosmos.esa.int/web/Gaia/dpac/consortium}). Funding for the DPAC
has been provided by National institutions, in particular the institutions
participating in the {\it Gaia} Multilateral Agreement.

This work has made use of the Python package GaiaXPy, developed and maintained by members of the Gaia Data Processing and Analysis Consortium (DPAC) and in  particular, Coordination Unit 5 (CU5), and the Data Processing Centre located at the Institute of Astronomy, Cambridge, UK (DPCI).

This work has made use of these python packages: pysynphot \citep{2013ascl.soft03023S},
synphot \citep{2018ascl.soft11001S},dustmaps \citep{2018JOSS....3..695M}, and extinction.

Guoshoujing Telescope (the Large Sky Area Multi-Object Fiber Spectroscopic Telescope LAMOST) is a National Major Scientific Project built by the Chinese Academy of Sciences. Funding for the project has been provided by the National Development and Reform Commission. LAMOST is operated and managed by the National Astronomical Observatories, Chinese Academy of Sciences.

Funding for the Sloan Digital Sky 
Survey IV has been provided by the 
Alfred P. Sloan Foundation, the U.S. 
Department of Energy Office of 
Science, and the Participating 
Institutions. 

SDSS-IV acknowledges support and 
resources from the Center for High 
Performance Computing  at the 
University of Utah. The SDSS 
website is www.sdss4.org.

SDSS-IV is managed by the 
Astrophysical Research Consortium 
for the Participating Institutions 
of the SDSS Collaboration including 
the Brazilian Participation Group, 
the Carnegie Institution for Science, 
Carnegie Mellon University, Center for 
Astrophysics | Harvard \& 
Smithsonian, the Chilean Participation 
Group, the French Participation Group, 
Instituto de Astrof\'isica de 
Canarias, The Johns Hopkins 
University, Kavli Institute for the 
Physics and Mathematics of the 
Universe (IPMU) / University of 
Tokyo, the Korean Participation Group, 
Lawrence Berkeley National Laboratory, 
Leibniz Institut f\"ur Astrophysik 
Potsdam (AIP),  Max-Planck-Institut 
f\"ur Astronomie (MPIA Heidelberg), 
Max-Planck-Institut f\"ur 
Astrophysik (MPA Garching), 
Max-Planck-Institut f\"ur 
Extraterrestrische Physik (MPE), 
National Astronomical Observatories of 
China, New Mexico State University, 
New York University, University of 
Notre Dame, Observat\'ario 
Nacional / MCTI, The Ohio State 
University, Pennsylvania State 
University, Shanghai 
Astronomical Observatory, United 
Kingdom Participation Group, 
Universidad Nacional Aut\'onoma 
de M\'exico, University of Arizona, 
University of Colorado Boulder, 
University of Oxford, University of 
Portsmouth, University of Utah, 
University of Virginia, University 
of Washington, University of 
Wisconsin, Vanderbilt University, 
and Yale University.

Data resources are supported by China National Astronomical Data Center (NADC) and Chinese Virtual Observatory (China-VO). This work is supported by Astronomical Big Data Joint Research Center, co-founded by National Astronomical Observatories, Chinese Academy of Sciences and Alibaba Cloud.

\bibliography{sample631}{}
\bibliographystyle{aasjournal}

\appendix

\section{Appendix A: Theoretical Basis of DDO51}

To evaluate the sensitivity of the DDO51 filter to stellar surface gravity and its effectiveness in measuring stellar atmospheric parameters when combined with other photometric bands, we analyzed its theoretical response to surface gravity variations using synthetic stellar spectra. The ATLAS9 stellar atmosphere models from \cite{castelli2003proc} and \cite{castelli2004new} were employed. This grid comprises 4, 300 theoretical stellar spectra spanning a wide range of metallicities, effective temperatures, and surface gravities. Specifically, the ATLAS9 grid includes models with metallicities of [M/H] = +0.5, +0.2, 0.0, -0.5, -1.0, -1.5, -2.0, and -2.5, and surface gravities ranging from log\,$g$ = 0.0 to +5.0 in steps of 0.5 dex. The effective temperature coverage extends from 3500 K to 50000 K, with a step size of 250 K within the 3750 K to 13000 K range. The spectra have a resolution of 20 \AA within the optical wavelength range of 0.29 - 1.00 µm\footnote{See details: https://www.stsci.edu/hst/instrumentation/reference-data-for-calibration-and-tools/astronomical-catalogs/castelli-and-kurucz-atlas.html}.

For our analysis, we selected ATLAS9 models with effective temperatures between 3750 K and 8000 K, utilizing the full range of surface gravities and metallicities available in the grid as described above. These models provide the stellar surface flux, $F_{\lambda}$, in units of $ergs\ cm^{-2}\ s^{-1} A^{-1}$.

Synthetic photometry involves convolving the absolute flux, $f_{\lambda}$, of an astronomical source with the system response function, $\eta_{\lambda}$, which accounts for the combined transmission of the telescope, filter, and detector. This convolution yields the integrated flux that would be measured by the specific photometric system, as described by the following equation:
\begin{equation}
\langle F_{\lambda}\rangle = \frac{\int\; F_{\lambda}\; \eta_{\lambda}\; \lambda\; d\lambda}{\int\; \eta_{\lambda}\; \lambda\; d\lambda}
\end{equation}

A thorough discussion of synthetic photometry can be found in \cite{casagrande2014synthetic}, \cite{bessell2012spectrophotometric}.

\begin{figure}[ht!]
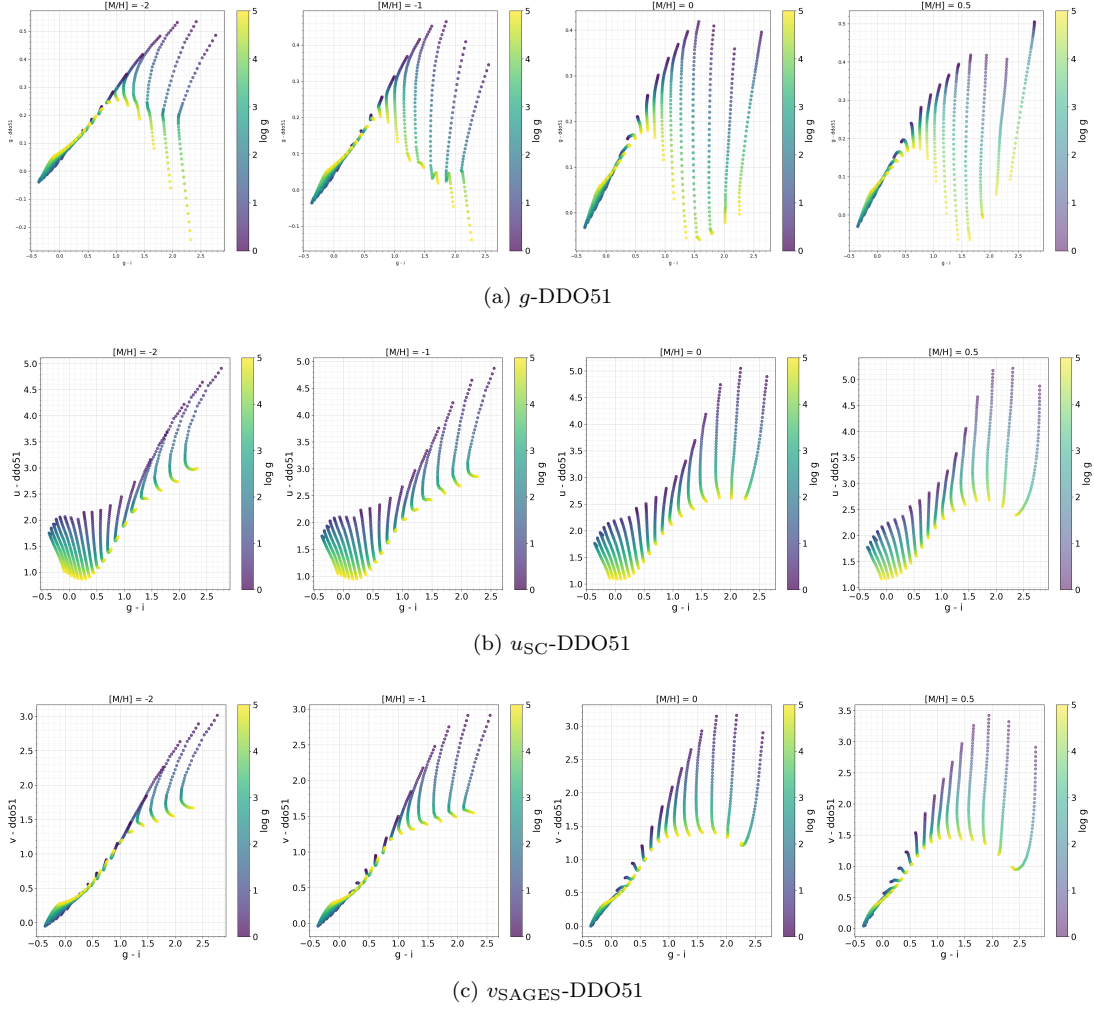

\centering
\gridline{\fig{g-DDO51.png}{0.8\textwidth}{(a) $g$-DDO51}
          }
          \gridline{\fig{u-DDO51.png}{0.8\textwidth}{(b) $u_{\rm SC}$-DDO51}}
\gridline{\fig{v-DDO51.png}{0.8\textwidth}{(c) $v_{\rm SAGES}$-DDO51}
          }

\caption{Color-color diagrams based on synthetic photometry from ATLAS9 model atmospheres, showing the behavior of DDO51, g, $u_{\rm SC}$, and $v_{\rm SAGES}$ colors as a function of metallicity and surface gravity. The $g - i$ color serves as a proxy for effective temperature. Panels (from top to bottom) show $g$-DDO51, $u_{\rm SC}$-DDO51, and $v_{\rm SAGES}$-DDO51 plotted against $g - i$. Point color indicates log\,$g$.}
\label{fig:kurucz_color}
\end{figure}

We constructed color-color diagrams using the u, v, g, and DDO51 bands for varying metallicities, as illustrated in Figure \ref{fig:kurucz_color}. The horizontal axis in each panel represents the g - i color, serving as a proxy for effective temperature (with higher temperatures corresponding to smaller color indices). As the figure depicts, a relatively continuous variation in g - DDO51 and v - DDO51 colors is observed for stars of different log\,$g$ values as temperature decreases, though this variation is modulated by metallicity. As log\,$g$ decreases, giant stars become progressively fainter in the DDO51 band relative to dwarf stars, creating a discernible separation in the color-color space. This enables a preliminary classification of stars as giants or dwarfs. However, refined surface gravity estimates require the incorporation of metallicity information, provided by the SAGES u and v filters. This motivates our subsequent use of a machine learning approach, leveraging the full SAGES photometric data set.

\section{Appendix B: Details of XGBoost model training}
XGBoost is a gradient boosting algorithm known for its robustness in handling non-linear relationships and feature interactions, widely used for predicting stellar parameters \citep[e.g.,][]{khalatyan2024transferring,rix2022poor,andrae2023robust}.

Key hyperparameters were tuned using Optuna framework, including 'n\_estimators', 'max\_depth', 'learning\_rate', 'subsample', 'colsample\_bytree', 'reg\_alpha', and 'reg\_lambda', optimizing for the Root Mean Squared Error (RMSE). Early stopping strategy was also used to prevent the model from over-fitting. 

Our tests demonstrated that XGBoost exhibits considerable robustness and is insensitive to the initial parameter settings, with similar prediction results obtained across different parameter configurations. The XGBoost parameters utilized were as follows: objective='reg:squarederror', eval\_metric='rmse', n\_estimators=500, max\_depth=8, and random\_state=42. 

\end{CJK*}
\end{document}